\documentclass[pre,aps,floats,twocolumn,superscriptaddress,floatfix,nofootinbib]
{revtex4}
\usepackage{graphicx,amssymb,amsmath,ifthen,rotating}
\usepackage[active]{srcltx}
\usepackage{color}
\begin{document}
\title{Self-gravitating clusters of Fermi-Dirac gas with planar, cylindrical, or spherical symmetry: evolution of density profiles with temperature}  
\author{Michael Kirejczyk}
\affiliation{
  Department of Physics,
  University of Rhode Island,
  Kingston RI 02881, USA}
  \author{Gerhard M{\"u}ller}
\affiliation{
  Department of Physics,
  University of Rhode Island,
  Kingston RI 02881, USA}
  \author{Pierre-Henri Chavanis}
\affiliation{
Laboratoire de Physique Th\'eorique, CNRS, 
Universit{\'e} Paul Sabatier, 
31062 Toulouse France}

\begin{abstract}
We calculate density profiles for self-gravitating clusters of an ideal Fermi-Dirac gas with nonrelativistic energy-momentum relation and macroscopic mass at thermal equilibrium.
Our study includes clusters with planar symmetry in dimensions $\mathcal{D}=1,2,3$, clusters with cylindrical symmetry in $\mathcal{D}=2,3$, and clusters with spherical symmetry in $\mathcal{D}=3$.
Wall confinement is imposed where needed for stability against escape.
The length scale and energy scale in use render all results independent of total mass and prove adequate at all temperatures.
We present exact analytic expressions for  (fully degenerate) $T=0$ density profiles in four of the six combinations of symmetry and dimensionality.
Our numerical results for $T>0$ describe the emergence, upon quasistatic cooling, of a core with incipient degeneracy surrounded by a more dilute halo.
The equilibrium macrostates are found to depend more strongly on the cluster symmetry than on the space dimensionality.
We demonstrate the mechanical and thermal stability of spherical clusters with coexisting phases.

\end{abstract}
\maketitle

%
\section{Introduction}\label{sec:intro}
%
The study of a self-gravitating Fermi-Dirac (FD) gas started in the context of white dwarf stars.
Fowler \cite{fowler} attributed their stability to the degenerate electron gas, whose quantum pressure arising from the Pauli exclusion principle balances the gravitational attraction at high density. 
Nonrelativistic white dwarfs at zero temperature represent a polytropic gas of index $n=3/2$ \cite{chandrabook}. 
The density profile, obtained by solving the  Lane-Emden equation \cite{emden}, has a compact support.
The fermion ball has a sharp surface at a finite radius.
Stoner \cite{stoner29}, Milne \cite{milne}, and Chandrasekhar \cite{chandra31nr} showed that the radius $r_0$ of the star decreases as the mass $m_\mathrm{tot}$ increases according to the law,
$m_\mathrm{tot}=91.9\, \hbar^6/(G^3m^8r_0^3)$ \cite{chandrabook}. 

Hertel and Thirring \cite{ht,htf} initiated the statistical mechanical analysis of FD clusters at nonzero temperature in the context of a study of nonrelativistic neutron stars.
They found density profiles with $\sim r^{-2}$ tails, which indicates that  finite-mass clusters at thermal equilibrium require artificial confinement.
Moreover, Hertel and Thirring \cite{htf} proved that the mean-field assumption, which neglects correlation effects, and the Thomas-Fermi approximation, which neglects some quantum effects, are highly accurate for macroscopic systems and exact in a specific thermodynamic limit. 

The work of Hertel and Thirring also brought forth evidence for the inequivalence of statistical ensembles in the face of long-range interactions, manifest e.g. in negative heat capacities.
This inequivalence is by no means unphysical. 
It locates the points of mechanical instability differently in clusters that are thermally isolated (the default in astrophysics) from clusters that are in contact with a heat bath (e.g. through the confining wall).
These (spinodal) points of instability are identified in caloric curves 
(inverse temperature versus negative internal energy) as points of infinite
slope (microcanonical ensemble) and points of zero slope (canonical ensemble).

Unlike the gravitational collapse of Maxwell-Boltzmann (MB) clusters
\cite{emden,antonov,lbw}, any precipitous contraction of FD clusters is arrested
by fermionic quantum statistics into a core-halo configuration.
Conversely, an increase in temperature (canonical ensemble) or internal energy (microcanonical ensemble) produces an abrupt change from a core-halo configuration to flat and more spread-out density profile.
The relevance of such behavior in the context of dark-matter research was
investigated by Bilic and Viollier \cite{bvn}.

Chavanis \cite{pt,ijmpb} conducted an exhaustive study of phase transitions in self-gravitating FD clusters using canonical and microcanonical ensembles, confirming the results and extending the work of Hertel and Thirring \cite{ht}.
Chavanis specifically identified and analyzed a zeroth-order phase transition in the microcanonical ensemble from gaseous to condensed macrostates, associated with a discontinuity in entropy.
The gaseous macrostate is located at the (spinodal) stability limit. 
He also discussed a first-order phase transition in the microcanonical ensemble between macrostates connected by a vertical Maxwell line on the caloric curve, associated with a discontinuity in temperature. 
This first-order phase transition does not take place in practice on account of the fact that the lifetimes of metastable states scale exponentially with the number of particles. \cite{lifetime}.

The zeroth-order phase transition has a counterpart at a different value of energy, from a condensed macrostate at the stability limit to a stable gaseous macrostate.
The two transitions are complementary to each other, one representing a mechanical instability on the way down in energy (collapse) and the other on the way up (explosion). 
In both cases the instability precipitates processes that are associated with an increase in entropy.

A similar scenario unfolds in the canonical ensemble, but at different landmarks on the caloric curves, the control parameter being now the temperature. 
In particular, the zeroth order phase transition is characterized by a discontinuity of free energy and the (unrealized) first-order phase transition is characterized by a discontinuity in energy associated with a horizontal Maxwell line.
Interestingly, Chavanis found that for tightly confined systems there is no phase transition, that for systems with intermediate confinement a phase transition takes place in the canonical ensemble but not in the microcanoncial ensemble, and for systems with loose confinement a phase transition takes place in both ensembles. 
This state of affairs is portrayed in intricate phase diagrams  \cite{ijmpb}. 
Similar results were obtained in the context of the fermionic King model \cite{clm2}, which does not require an artificial confinement.

Further light on the phase behavior of self-gravitating FD clusters has been shed by studies that extended the space dimensionality away from $D=3$ to lower and higher values.
They include Refs.~\cite{Chav07} and \cite{ptdimd} for
properties at
$T=0$ and $T>0$, respectively.
One important insight is that $D=4$ is a sort of upper marginal dimensionality
for the mechanical stability of nonrelativistic FD clusters against
gravitational collapse.\footnote{For $D\ge 4$, quantum mechanics cannot
stabilize matter against gravitational collapse even in the 
nonrelativistic regime \cite{Chav07,ptdimd}. This is similar to a result found
by Ehrenfest \cite{ehrenfest} who considered the effect of the dimension of
space on the laws of physics and showed that planetary motion and the Bohr atom
would not be stable in a space of dimension $D\ge 4$.} 
A second insight is that the phase behavior exhibits features of universality across ranges of dimensionality. 

In addition to white dwarf and neutron star research \cite{shapiroteukolsky}, the physics of FD clusters has applications in the study of conjectured dark matter halos made of massive neutrinos. 
As discussed in \cite{clm2}, the FD distribution function may be justified by the theory of violent relaxation of Lynden-Bell \cite{lb}.\footnote{The statistical equilibrium state of a dark matter halo may result from a process of collisionless relaxation \cite{lb,chavmnras} rather than from a process of collisional relaxation because the relaxation time due to two-body gravitational encounters in dark matter halos is much larger than the age of the universe. However, the collisional relaxation time may be reduced if the fermions are self-interacting.}
The equilibrium states have a core-halo structure made of a quantum core (fermion ball) surrounded by an isothermal halo (envelope) \cite{csmnras}. 
Some authors \cite{bmtv,btv,rar,rarnew} have proposed that a very compact quantum core  made of dark matter fermions with mass $\sim50\,{\rm keV/c^2}$ could mimic a supermassive black hole at the center of the galaxies.
Other authors \cite{clm2,mcmh} have considered a smaller fermion mass of
$\sim1\, {\rm keV}/c^2$ and argued that the quantum core has 
the shape of a large dark matter bulge (see \cite{modeldm}
for a
comparison between these two scenarios). 
The nature of dark matter, let alone the values of particle mass, are very much in dispute.

This work, which emphasizes the distinction between cluster symmetry and space dimensionality, begins with establishing the conditions of mechanical stability and thermal equilibrium (Sec.~\ref{sec:equi}).
A second point of emphasis is the choice of practical scales for length and energy (Sec.~\ref{sec:scales}).
In the discussion of fully degenerate finite-mass clusters the emphasis is on exact results for density profiles (Sec.~\ref{sec:degen}).
The centerpieces for the analysis and interpretation of FD clusters at $T>0$ are robust free-energy expressions and caloric curves (Sec.~\ref{sec:caloric}).
With these tools in place we are ready to analyze density profiles subject
to gradual and abrupt changes upon cooling or heating (Sec.~\ref{sec:emerge})
and to describe phase-coexisting macrostates (Sec.~\ref{sec:phase coex}).

%
\section{Equilibrium Conditions}\label{sec:equi}
%
The density profiles of self-gravitating FD gas clusters analyzed in this work are governed by mechanical stability and thermal equilibrium. 
The former is encoded in an equation of motion (EOM) and the latter in an equation of state (EOS).
In the present context, mechanical stability means hydrostatic equilibrium.
The only particle interaction included is the mutual gravitational attraction. The mean-field assumption is validated by the long range of this force \cite{htf}.

We consider clusters with planar symmetry $(\mathcal{D}_\sigma=1)$, cylindrical symmetry $(\mathcal{D}_\sigma=2)$, and spherical symmetry $(\mathcal{D}_\sigma=3)$.
All profiles are functions of the distance $r$ from the center of the cluster.
For $\mathcal{D}_\sigma=1$, the center is a point, a line, or a plane in $\mathcal{D}=1,2,3$, respectively. 
For $\mathcal{D}_\sigma=2$, the center is a point or a line in $\mathcal{D}=2,3$, respectively. 
For $\mathcal{D}_\sigma=3$, the center is a point (in ${\mathcal{D}=3}$). 
We thus write $\rho_\mathrm{v}(r)$ and $p(r)$ for the radial profiles of particle density and pressure, respectively.
The temperature $T$ is uniform.

The total number of particles in a finite cluster is obtained from the density profile via the integral,
\begin{equation}\label{eq:4} 
N=L^{\mathcal{D}-\mathcal{D}_\sigma}
\int_0^R dr\,\mathcal{A}_{\mathcal{D}_\sigma} 
r^{\mathcal{D}_\sigma-1}\rho_\mathrm{v}(r),
\end{equation}
where $R$ is the radius of the confining wall, $L$ the length of a cylinder or
of the sides of a plane in cases with $\mathcal{D}_\sigma<\mathcal{D}$, and
$\mathcal{A}_\mathcal{D}\doteq2\pi^{\mathcal{D}/2}/\Gamma(\mathcal{D}/2)$ is the
surface area of the $\mathcal{D}$-dimensional unit sphere.
The condition $L\gg R$ guarantees that deviations from the symmetry assumed to hold are negligible.

Hydrostatic equilibrium relates pressure and density,
\begin{equation}\label{eq:6} 
\frac{d}{dr}p(r)=M\rho_\mathrm{v}(r)g(r),
\end{equation}
involving the gravitational field,
\begin{equation}\label{eq:8} 
g(r)=-\frac{\mathcal{A}_\mathcal{D}G_\mathcal{D}M}{r^{\mathcal{D}_\sigma-1}}
\int_0^r dr'r'^{\mathcal{D}_\sigma-1}\rho_\mathrm{v}(r').
\end{equation}
In order to accommodate scenarios which include plasmas, we distinguish the kinetic mass $m$ (e.g. of electrons) entering the EOS and the gravitational mass $M$ (e.g. of nucleons per electron) entering the EOM. The total mass of a cluster is denoted $m_\mathrm{tot}$.
The strength $G_\mathcal{D}$ of the gravitational interaction is empirically known, of course, only in $\mathcal{D}=3$.

The EOS (in the local density approximation) for a nonrelativistic FD gas in $\mathcal{D}$ dimensions is implicit in the relations, 
\begin{subequations}\label{eq:1}
\begin{align}\label{eq:1b}
& \rho_\mathrm{v}(r)\lambda_T^\mathcal{D}=
g_s\,f_{\mathcal{D}/2}\big(z(r)\big), \\ 
\label{eq:1c}
& u_\mathrm{v}(r)\lambda_T^\mathcal{D}
=\frac{\mathcal{D}}{2}k_\mathrm{B}Tg_s\,
f_{\mathcal{D}/2+1}\big(z(r)\big), \\
\label{eq:1a}
& \frac{p'(r)\lambda_T^\mathcal{D}}{k_\mathrm{B}T}
=g_s\,\frac{d}{dz}f_{\mathcal{D}/2+1}(z)z'(r), 
\end{align}
\end{subequations}
between the particle density, the kinetic energy density, and the pressure, parametrized by the fugacity profile $z(r)$.
Here $g_s$ is the spin degeneracy, $\lambda_T=\sqrt{h^2/2\pi mk_\mathrm{B}T}$ is the de Broglie thermal wavelength, and 
\begin{equation}\label{eq:3}
f_n(z)\doteq\frac{1}{\Gamma(n)}\int_0^\infty\frac{dx\,x^{n-1}}{z^{-1}e^x+1}
=\sum_{l=1}^\infty(-1)^{l-1}\frac{z^l}{l^n},\quad z\geq0
\end{equation}
are the (polylogarithmic) FD functions with the familiar special cases, $f_0(z)=z/(1+z)$, $f_1(z)=\ln(1+z)$, and $f_\infty(z)=z$.
The entropy density, 
\begin{equation}\label{eq:74} 
\frac{s\lambda_T^\mathcal{D}}{g_sk_\mathrm{B}}=
\left(\frac{\mathcal{D}}{2}+1\right)f_{\mathcal{D}/2+1}(z)
-\ln z\,f_{\mathcal{D}/2}(z).
\end{equation}
is inferred from (\ref{eq:1}) via Euler's equation.

We have rendered Eq.~(\ref{eq:1a}) in a way that consistency with (\ref{eq:6}) is ensured even under phase coexistence. 
Substitution of (\ref{eq:1a}) into (\ref{eq:6}) with use of (\ref{eq:8}) and the recurrence relation $zf'_n(z)=f_{n-1}(z)$ produces the ODE, 
\begin{equation}\label{eq:11} 
\frac{z''}{z}+\frac{\mathcal{D}_\sigma-1}{r}\frac{z'}{z}-\left(\frac{z'}{z}\right)^2
+\frac{g_s\mathcal{A_D}G_\mathcal{D}M^2}{\lambda_T^\mathcal{D}k_\mathrm{B}T}f_{\mathcal{D}/2}(z)=0,
\end{equation}
for the fugacity profile, from which the density profile follows via (\ref{eq:1b}).

For (thermodynamically) open systems, which include clusters of finite and infinite mass, the boundary conditions are
\begin{equation}\label{eq:12} 
z'(0)=0,\quad 0<z(0)=z_0,
\end{equation}
with the (average) total mass, $m_\mathrm{tot}=NM$, provided it is finite, inferred from (\ref{eq:4}).
Closed systems of finite mass may only exist under confinement such as imposed by a wall at $R<\infty$.
For cases with $\mathcal{D}_\sigma<\mathcal{D}$, it is useful to introduce a
rescaled number of particles: 
\begin{equation}\label{eq:13}
\tilde{N}\doteq\frac{N}{L^{\mathcal{D}-\mathcal{D}_\sigma}}.
\end{equation}
The second boundary condition (\ref{eq:12}) is to be replaced (for closed systems) by the integral (\ref{eq:4}) converted into
\begin{equation}\label{eq:14}
\frac{g_s\mathcal{A}_{\mathcal{D}_\sigma}}{\tilde{N}\lambda_T^{\mathcal{D}}}
\int_0^Rdr\,r^{\mathcal{D}_\sigma-1}f_{\mathcal{D}/2}(z)=1.
\end{equation}

%
\section{Scaling Convention}\label{sec:scales}
%
For our analysis we have constructed a length scale and an energy scale that are useful for the description of macroscopic FD clusters at all temperatures including $T=0$.
These scales work equally well for the gaseous part of Bose-Einstein (BE)
clusters \cite{sgcbe}.
The length scale $r_\mathrm{s}$ and temperature scale $T_\mathrm{s}$ are derived from the thermal wavelength and a macroscopic reference volume as follows:
\begin{equation}\label{eq:15} 
\hat{r}\doteq\frac{r}{r_\mathrm{s}},\quad \hat{T}\doteq\frac{T}{T_\mathrm{s}},
\end{equation}
\begin{equation}\label{eq:17} 
\tilde{N}\lambda_{T_\mathrm{s}}^{\mathcal{D}}
=\frac{\mathcal{A}_{\mathcal{D}_\sigma}}{\mathcal{D}_\sigma}\,r_\mathrm{s}^{\mathcal{D}_\sigma},
\quad
\frac{1}{r_\mathrm{s}^2}=\frac{1}{2\mathcal{D}_\sigma}\frac{\mathcal{A_D}G_\mathcal{D}M^2}
{\lambda_{T_\mathrm{s}}^{\mathcal{D}}k_\mathrm{B}T_\mathrm{s}}.
\end{equation}
Relations (\ref{eq:17})
determine $r_\mathrm{s}$ and $T_\mathrm{s}$ as functions of particle mass $m$ and total mass $m_\mathrm{tot}=NM$ for cases with $\mathcal{D}_\sigma=\mathcal{D}$ (see Appendix~\ref{sec:appa}).
If $\mathcal{D}_\sigma<\mathcal{D}$ we use $\tilde{m}_\mathrm{tot}=\tilde{N}M$
with $\tilde{N}$ from (\ref{eq:13}) instead, which can still be used as a measure for how massive the cluster is.

Expressed in the dimensionless variables thus defined, including the relation $\hat{z}(\hat{r})\doteq z(r)$, the ODE (\ref{eq:11}) becomes
\begin{equation}\label{eq:23} 
\frac{\hat{z}''}{\hat{z}}+\frac{\mathcal{D}_\sigma-1}{\hat{r}}\frac{\hat{z}'}{\hat{z}}
-\left(\frac{\hat{z}'}{\hat{z}}\right)^2
+\frac{2\mathcal{D}_\sigma}{\hat{T}}\rho(\hat{r})=0,
\end{equation}
where we use the dimensionless density,
\begin{equation}\label{eq:22}
\rho(\hat{r})\doteq \lambda_{T_\mathrm{s}}^\mathcal{D}\rho_\mathrm{v}(\hat{r})
=g_s\,\hat{T}^{\mathcal{D}/2}f_{\mathcal{D}/2}(\hat{z}),
\end{equation}
and rewrite (\ref{eq:14}) in the form,
\begin{equation}\label{eq:24}
\mathcal{D}_\sigma \int_0^{\hat{R}} d\hat{r}\,\hat{r}^{\mathcal{D}_\sigma-1}
\rho(\hat{r})=1.
\end{equation}
Contact with a frequently used alternative scaling convention, whose advantage is to demonstrate the classical limit, is established in Appendix~\ref{sec:appa}.

%
\section{Degenerate clusters}\label{sec:degen}
%
FD clusters of finite mass at zero temperature are stable against particle escape.
They are also stable against gravitational collapse for $1 \leq \mathcal{D}_\sigma \leq \mathcal{D} =3$ as long as the total mass does not push the Fermi momentum into the relativistic regime.
Density profiles of fully degenerate FD clusters are important anchor points for the analysis of the effects of rising temperature (Sec.~\ref{sec:emerge}).
The effects of combinations of temperature and mass variations will be investigated elsewhere \cite{sgcrfd}.

For the purpose of investigating  $T=0$ density profiles, we distill out of the ODE (\ref{eq:11}) for the fugacity $z(r)$ an ODE for the chemical potential $\mu(r)=k_\mathrm{B}T\ln z(r)$ in the limit $T\to0$,
using the leading term in the asymptotic expansion of FD functions,
\begin{equation}\label{eq:40}  
f_{\mathcal{D}/2}(z)\leadsto
\frac{(\beta\mu)^{\mathcal{D}/2}}{\Gamma(\mathcal{D}/2+1)}.
\end{equation}
The ODE for the chemical potential, expressed in scaled variables, $\hat{\mu}\doteq\mu/k_\mathrm{B}T_\mathrm{s}$ and ${\hat{r}\doteq r/r_\mathrm{s}}$, which for $\mathcal{D}_\sigma=\mathcal{D}$ has the structure of a Lane-Emden equation, reads
\begin{equation}\label{eq:42} 
\hat{\mu}''+\frac{\mathcal{D}_\sigma-1}{\hat{r}}\hat{\mu}'
+\frac{2g_s\mathcal{D}_\sigma}{\Gamma(\mathcal{D}/2+1)}\,\hat{\mu}^{\mathcal{D}/2}=0.
\end{equation}
We are seeking a solution with boundary conditions, 
\begin{align}\label{eq:176a}
& \hat{\mu}'(0)=0, \quad \hat{\mu}(\hat{r}_0)=0,
\end{align}
where the cluster radius $\hat{r}_0$ is implicitly determined by the normalization condition (\ref{eq:24}) adapted as follows:
\begin{align}\label{eq:176b}
\frac{g_s\mathcal{D}_\sigma}{\Gamma(\mathcal{D}/2+1)}\int_0^{\hat{r}_0}
d\hat{r}\,\hat{r}^{\mathcal{D}_\sigma-1}
\big[\hat{\mu}(\hat{r})\big]^{\mathcal{D}/2}=1.
\end{align}
The tacit assumption is that $\hat{R}>\hat{r}_0$.
Fully  degenerate FD clusters are self-confined.

The (scaled) density and pressure profiles, expressed via chemical potential, inferred from Eqs.~(\ref{eq:1}) and the asymptotics (\ref{eq:40}), become
\begin{equation}\label{eq:44}
\rho(\hat{r})=\frac{g_s\big[\hat{\mu}(\hat{r})\big]^{\mathcal{D}/2}}{\Gamma(\mathcal{D}/2+1)}
\theta(\hat{r}_0-\hat{r}),
\end{equation}
\begin{equation}\label{eq:43} 
\hat{p}(\hat{r})\doteq\frac{p(r)}{k_\mathrm{B}T_\mathrm{s}\lambda_{T_\mathrm{s}}^{-\mathcal{D}}}
=\frac{g_s\big[\hat{\mu}(\hat{r})\big]^{\mathcal{D}/2+1}}{\Gamma(\mathcal{D}/2+2)}
\theta(\hat{r}_0-\hat{r}),
\end{equation}
which confirms the result of \cite{Chav07} that we are dealing with polytropes of index $n=\mathcal{D}/2$, independent of $\mathcal{D}_\sigma$ (see Appendix~\ref{sec:appd} for more details).
The linear cusp of $\hat{\mu}(\hat{r})$ at $\hat{r}_0$ determines the power-law cusp singularities of the density $\rho(\hat{r})$ and the pressure $\hat{p}(\hat{r})$ at the surface of the cluster via (\ref{eq:44}) and (\ref{eq:43}).

In the following we analyze the solutions of Eqs.~(\ref{eq:42})-(\ref{eq:176b}) separately for the six combinations $0\leq\mathcal{D}_\sigma\leq\mathcal{D}\leq3$.
Within the nonrelativistic regime, all scaled profiles are universal, i.e. independent of total mass.

\subsection{$\mathcal{D}_\sigma=\mathcal{D}=1$}\label{sec:degen-planar-d1}
The ODE (\ref{eq:42}) for the scaled chemical potential $\hat{\mu}(\hat{r})$ simplifies into
\begin{equation}\label{eq:180} 
\hat{\mu}''+\frac{2\,g_s}{\Gamma(3/2)}\hat{\mu}^{1/2}=0.
\end{equation}
We solve it by transcribing it to
\begin{equation}\label{eq:181} 
\hat{r}''-\frac{2\,g_s}{\Gamma(3/2)}\sqrt{\hat{\mu}}\,\big[\hat{r}'\big]^3=0,
\end{equation}
for the inverse function $\hat{r}(\hat{\mu})$, thus reducing it effectively to first order (see Ref.~\cite{CS04} for a different approach).
The first integral is carried out by separation of the variables $\hat{\mu}$ and $\hat{s}\doteq\hat{r}'(\hat{\mu})$ with use of the boundary condition, $\hat{r}'(\hat{\mu}_0)=-\infty$:
\begin{equation}\label{eq:183}
 \int_{-\infty}^{\hat{s}}\frac{d\hat{s}'}{\hat{s}'^3}=\frac{2\,g_s}{\Gamma(3/2)}
 \int_{\hat{\mu}_0}^{\hat{\mu}}d\hat{\mu}'\sqrt{\hat{\mu}'}.
\end{equation}
We thus obtain
\begin{equation}\label{eq:184} 
\hat{r}'(\hat{\mu})=-\left(\frac{3\,\Gamma(3/2)}{8\,g_s}\right)^{1/2}\sqrt{\frac{1}{\hat{\mu}_0^{3/2}-\hat{\mu}^{3/2}}}.
\end{equation}
The inverse profile then reads
\begin{equation}\label{eq:185}
 \hat{r}(\hat{\mu})=\hat{r}_0-\left(\frac{3\,\Gamma(3/2)}{8\,g_s}\right)^{1/2}\int_0^{\hat{\mu}}\frac{d\hat{\mu}'}{\sqrt{\hat{\mu}_0^{3/2}-\hat{\mu}^{3/2}}},
 \end{equation}
 for $0\leq\hat{\mu}\leq\hat{\mu}_0$.
 Implementing $\hat{r}(\hat{\mu}_0)=0$ from (\ref{eq:176a}) yields
\begin{align}\label{eq:186}
 \hat{r}_0 &=\left(\frac{3\,\Gamma(3/2)}{8\,g_s}\right)^{1/2}\hat{\mu}_0^{1/4}\int_0^1\frac{dx}{\sqrt{1-x^{3/2}}} \\
 &=\frac{1}{4}\sqrt{\frac{3\pi^{3/2}}{g_s}}\frac{\Gamma(5/3)}{\Gamma(7/6)}\,\hat{\mu}_0^{1/4}.
\end{align}
Condition (\ref{eq:176b}) is satisfied by the values
${\hat{\mu}_0=0.76162}$ and $\hat{r}_0=0.656793$ if we set $g_s=2$.

\begin{figure}[htb]
  \begin{center}
 \includegraphics[width=.65 \linewidth]{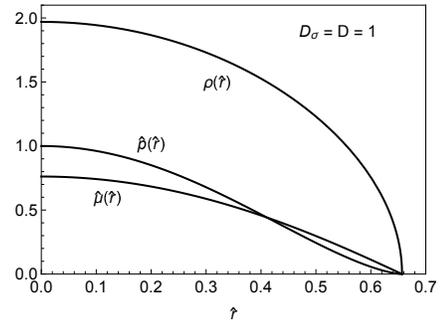}  
\end{center}
\caption{Universal profiles for the scaled chemical potential, density, and pressure of the nonrelativistic FD gas with $\mathcal{D}_\sigma=\mathcal{D}=1$ at $T=0$.}
  \label{fig:1}
\end{figure}

In Fig.~\ref{fig:1} we plot the profile for the chemical potential derived from the solution (\ref{eq:185}) and the profiles (\ref{eq:44}) and (\ref{eq:43}) 
for density and pressure, respectively.
The function $\hat{\mu}(\hat{r})$ vanishes linearly at $\hat{r}=\hat{r}_0$ (to
leading order), implying cusp singularities, $\rho\sim(\hat{r}_0-\hat{r})^{1/2}$
and $\hat{p}\sim(\hat{r}_0-\hat{r})^{3/2}$ for the other two functions.
When particles are added i.e. when the total mass increases, the radius of the cluster increases at the rate $r_0\sim N^{1/3}$ and the pressure at the center of the cluster increases at the rate $p(0)\sim N^2$.

\subsection{$\mathcal{D}_\sigma=1$, $\mathcal{D}=2$}\label{sec:degen-planar-d2}
\begin{figure}[htb]
  \begin{center}
 \includegraphics[width=0.65\linewidth]{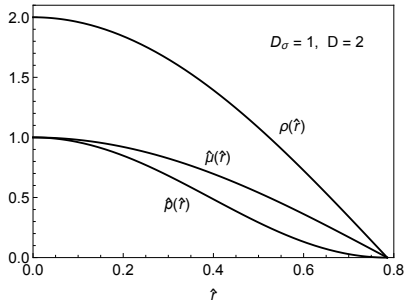}  
\end{center}
\caption{Universal profiles for the scaled chemical potential, density, and pressure of the nonrelativistic FD gas with $\mathcal{D}_\sigma=1$, $\mathcal{D}=2$ at $T=0$.}
  \label{fig:2}
\end{figure}

For this case the ODE (\ref{eq:42}) becomes linear,
\begin{equation}\label{eq:189} 
\hat{\mu}''+2\,g_s\hat{\mu}=0
\end{equation}
 and the solution for $g_s=2$ which satisfies the boundary conditions (\ref{eq:176a})-(\ref{eq:176b}) takes the form,
\begin{equation}\label{eq:190}
 \hat{\mu}(\hat{r})=\cos(2\,\hat{r}),
\end{equation}
implying $\hat{r}_0=\pi/4$ and $\hat{\mu}_0=1$.
This profile along with the profiles (\ref{eq:44}) and (\ref{eq:43}) 
are shown in Fig.~\ref{fig:2}.

The chemical potential and the density have proportional profiles, approaching zero linearly at $\hat{r}_0$, whereas the pressure vanishes quadratically at the surface of the cluster.
Adding particles does not change the radius of the cluster.
In consequence, the local density grows linearly with mass.
The central pressure increases quadratically with mass.

\subsection{$\mathcal{D}_\sigma=1,~ \mathcal{D}=3$}\label{sec:degen-planar-d3}
\begin{figure}[htb]
  \begin{center}
 \includegraphics[width=0.65\linewidth]{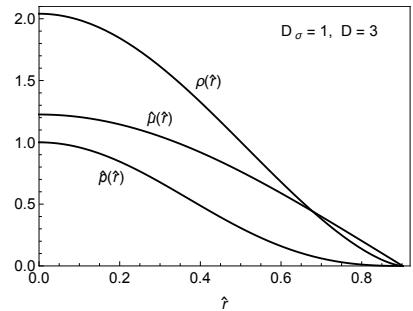}  
\end{center}
\caption{Universal profiles for the scaled chemical potential, density, and
pressure of the 
nonrelativistic FD gas with $\mathcal{D}_\sigma=1$, $\mathcal{D}=3$ at $T=0$.}
  \label{fig:3}
\end{figure}

The ODE to be solved in this case is
\begin{equation}\label{eq:193} 
\hat{\mu}''+\frac{2\,g_s}{\Gamma(5/2)}\hat{\mu}^{3/2}=0
\end{equation}
for the function $\hat{\mu}(\hat{r})$ or the (effectively first-order) ODE,
\begin{equation}\label{eq:194} 
\hat{r}''-\frac{2\,g_s}{\Gamma(5/2)}\hat{\mu}^{3/2}\,\big[\hat{r}'\big]^3=0,
\end{equation}
for the inverse function $\hat{r}(\hat{\mu})$.
The resulting inverse profile becomes
\begin{equation}\label{eq:195}
 \hat{r}(\hat{\mu})=\hat{r}_0-\left( \frac{5\, \Gamma(5/2)}{8\, g_s} \right)^{1/2}\int_0^{\hat{\mu}}
 \frac{d\hat{\mu}'}{\sqrt{\hat{\mu}_0^{5/2}-\hat{\mu}'^{5/2}}},
\end{equation}
 for $0\leq\hat{\mu}\leq\hat{\mu}_0$.
 Implementing conditions (\ref{eq:176a})-(\ref{eq:176b}) yields the relation 
\begin{align}\label{eq:196}
 \hat{r}_0 &=\left( \frac{5\, \Gamma(5/2)}{8\, g_s} \right)^{1/2}\hat{\mu}_0^{-1/4}\int_0^1\frac{dx}{\sqrt{1-x^{5/2}}} \nonumber \\
 &=\frac{1}{4}\sqrt{\frac{15 \pi^{3/2}}{2\, g_s}}\frac{\Gamma(7/5)}{\Gamma(9/10)}\,\hat{\mu}_0^{-1/4},
\end{align}
and the values $\hat{\mu}_0=1.225233$, $\hat{r}_0=0.901549$.
The  profiles (\ref{eq:44}) and (\ref{eq:43}) 
follow directly (see Fig.~\ref{fig:3}).

The linear cusp of the chemical potential at $\hat{r}_0$ is universal in $\mathcal{D}_\sigma=1$. 
The cusp singularities of density and pressure become weaker as $\mathcal{D}$ increases.
The (unscaled) radius $r_0$ shrinks with increasing mass: $r_0\sim N^{-1/5}$. 

\subsection{$\mathcal{D}_\sigma=\mathcal{D}=2$}\label{sec:degen-cylind-d2}
The term with a first-order derivative in the  ODE (\ref{eq:42}) for $\mathcal{D}_\sigma>1$ removes the advantage of switching to inverse functions.
However, for $\mathcal{D}=2$, the ODE,
\begin{equation}\label{eq:199} 
\hat{\mu}''+\frac{1}{\hat{r}}\,\hat{\mu}'+4\,g_s\hat{\mu}=0.
\end{equation}
is recognizable as characterizing Bessel functions. 
The solution which satisfies the boundary conditions (\ref{eq:176a})-(\ref{eq:176b}) with $g_s=2$ is well known \cite{Chav07}:
\begin{equation}\label{eq:200}
\hat{\mu}(\hat{r})=\hat{\mu}_0\,{J}_0(2\sqrt{2}\hat{r}).
\end{equation}
\begin{figure}[b!]
  \begin{center}
 \includegraphics[width=.65\linewidth]{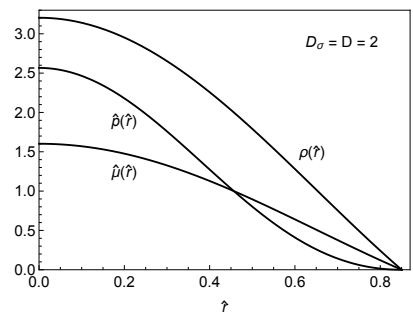}  
\end{center}
\caption{Universal profiles for the scaled chemical potential, density, and pressure of the nonrelativistic FD gas with $\mathcal{D}_\sigma=\mathcal{D}=2$ at $T=0$.}
  \label{fig:4}
\end{figure}
The radius $\hat{r}_0$ of the cluster is determined by the first zero of the Bessel function.
The value $\hat{\mu}_0$ then follows from (\ref{eq:176b}).
We thus obtain $\hat{\mu}_0=1.60198$ and $\hat{r}_0=0.850234$.
The profile (\ref{eq:200}) along with the profiles (\ref{eq:44}) and (\ref{eq:43})
are shown in Fig.~\ref{fig:4}. 

The chemical potential and the density are directly proportional and vanish linearly at the edge of the cluster whereas the pressure vanishes quadratically.
Features shared by all cases with $\mathcal{D}=2$ include that the average density and total mass are proportional, the radius of the cluster is independent of the mass, and the central pressure increases quadratically with mass.

\subsection{$\mathcal{D}_\sigma=2,~ \mathcal{D}=3$}\label{sec:degen-cylind-d3}
\begin{figure}[htb]
  \begin{center}
 \includegraphics[width=.60\linewidth]{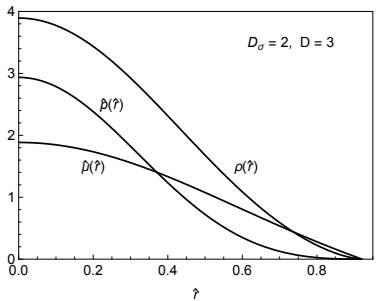}  
\end{center}
\caption{Universal profiles for the scaled chemical potential, density, and pressure of the nonrelativistic FD gas with $\mathcal{D}_\sigma=2$, $\mathcal{D}=3$ at $T=0$.}
  \label{fig:5}
\end{figure}

This case requires that we numerically solve the ODE
\begin{equation}\label{eq:203} 
\hat{\mu}''+\frac{1}{\hat{r}}\,\hat{\mu}'+\frac{4\,g_s}{\Gamma(5/2)}\hat{\mu}^{3/2}=0,
\end{equation}
subject to the three simultaneous conditions (\ref{eq:176a})-(\ref{eq:176b}).
The solution is unique and has the values, $\hat{\mu}_0=1.88488$, for the central fugacity and, $\hat{r}_0=0.92098$, for the cluster radius.
The solution $\hat{\mu}(\hat{r})$ of (\ref{eq:203}) and the profiles of $\rho(\hat{r})$, $\hat{p}(\hat{r})$ inferred from (\ref{eq:44}), (\ref{eq:43}) are shown in Fig.~\ref{fig:5}.

\subsection{$\mathcal{D}_\sigma=3,~ \mathcal{D}=3$}\label{sec:degen-sphere-d3}
\begin{figure}[htb]
  \begin{center}
 \includegraphics[width=.65\linewidth]{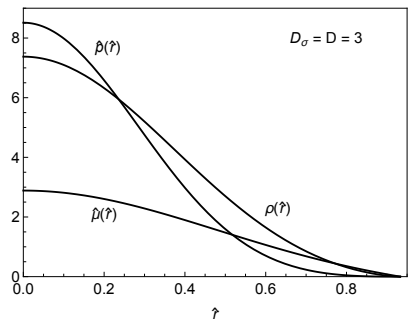}  
\end{center}
\caption{Universal profiles for the scaled chemical potential, density, 
and pressure of the nonrelativistic FD gas in $\mathcal{D}_\sigma=\mathcal{D}=3$
at $T=0$.}
  \label{fig:6}
\end{figure}

The universal profiles for this case are of textbook familiarity
\cite{chandrabook}.
The analysis of 
\begin{equation}\label{eq:206} 
\hat{\mu}''+\frac{2}{\hat{r}}\,\hat{\mu}'+\frac{6\,g_s}{\Gamma(5/2)}\hat{\mu}^{3/2}=0.
\end{equation}
must again be carried out numerically in its entirety.
The specifications of the solution which satisfies the three conditions (\ref{eq:176a})-(\ref{eq:176b}) are $\hat{\mu}_0=2.88587$ and $\hat{r}_0=0.932973$.
The resulting profiles of $\hat{\mu}(\hat{r})$, $\rho(\hat{r})$, and $\hat{p}(\hat{r})$ are shown in Fig.~\ref{fig:6}.
While the radius shrinks in real space when particles are added, $r_0\sim N^{-1/3}$, it grows in reciprocal space, $\mu_0\sim N^{4/3}$.
The central pressure rises rapidly with increasing mass: $p(0)\sim N^{10/3}$. 

\subsection{Mass-radius relation}\label{sec:mass-radius}
Our choice of length scale has the advantage of producing universal profiles. 
Clusters of given symmetry and dimensionality have their surface at a specific numerical value of the scaled radius $\hat{r}_0$ irrespective of mass (within the nonrelativistic regime).
In consequence, the mass-radius relation of self-gravitating clusters -- a quantity of considerable interest -- is encoded in the length scale $r_\mathrm{s}$.

In Table~\ref{tab:1} we state the explicit dependence of $r_\mathrm{s}$ on the number of particles $N$, and on the relevant kinetic mass $m$ and gravitational mass $M$ of particles.
The total mass is $m_\mathrm{tot}=NM$.
In ordinary matter $m$ is the electron mass and $M$ the mass of nucleons per electron.

\begin{table}[htb]
\caption{Explicit dependence of the length scale $r_\mathrm{s}$, which determines the cluster radius at $T=0$ via $r_0=\hat{r}_0r_\mathrm{s}$ on the number $N$ of particles with (kinetic) mass $m$ and (gravitational) mass $M$. The total mass is $m_\mathrm{tot}=NM$. For cases with $\mathcal{D}_\sigma<\mathcal{D}$ we use $\tilde{N}$ as defined in (\ref{eq:13}). The six relations are extracted from (\ref{eq:17}).}\label{tab:1}
\begin{center}
\begin{tabular}{cc|l}\rule[-2mm]{0mm}{6mm}
$\mathcal{D}_\sigma$~ & ~$\mathcal{D}$~~ & \hspace{20mm}$r_\mathrm{s}$  \\ \hline \rule[-2mm]{0mm}{9mm}
1 & 1 & $\displaystyle 
~~ \left(\frac{\pi\hbar^2}{G_1}\right)^{1/3}
\frac{N^{1/3}}{M^{2/3}m^{1/3}}$ \\ \rule[-2mm]{0mm}{9mm}
1 & 2 & $\displaystyle 
~~ \hbar\sqrt{\frac{2}{G_2}}M^{-1}m^{-1/2}$ \\ 
\rule[-2mm]{0mm}{9mm}
1 & 3 & $\displaystyle 
~~ \left(\frac{2\hbar^6}{G_3^3}\right)^{1/5}\tilde{N}^{-1/5}M^{-6/5}m^{-3/5}$ \\ \rule[-2mm]{0mm}{9mm}
2 & 2 & $\displaystyle 
~~ \frac{2\hbar}{\sqrt{G_2}}M^{-1}m^{-1/2}$ \\
\rule[-2mm]{0mm}{9mm}
2 & 3 & $\displaystyle
~~ \left(\frac{8\pi\hbar^6}{G_3^3}\right)^{1/4}\tilde{N}^{-1/4}M^{-3/2}m^{-3/4}$ \\ \rule[-2mm]{0mm}{9mm}
3 & 3 & $\displaystyle
~~ (36\pi)^{1/3}\frac{\hbar^2}{G_3}N^{-1/3}M^{-2}m^{-1}$
\end{tabular}
\end{center}
\end{table}

The relations compiled in Table~\ref{tab:1}, at least the entries for
$\mathcal{D}=3$, are likely to be of importance 
in dark matter research, where it is still unclear what particle mass $M$ might
dominate gravity and what particle mass $m$ might dominate the pressure (via the
thermal wavelength).
The distinct dependences of the cluster radius on the two particle masses is likely  to yield clues about the range of particle masses that qualify as constituents of dark matter. 

%
\section{Free energies and Caloric curves}\label{sec:caloric}
%
For the analysis of density profiles at $T>0$ it is useful to have caloric curves -- functional relations between inverse scaled temperature and negative scaled internal energy -- available as a road map on which relevant landmarks can be identified, namely points of instability in both the canonical and microcanonical ensembles.
Caloric curves were analyzed in Ref.~\cite{ptdimd} for $\mathcal{D}=\mathcal{D}_\sigma$ using a different scaling convention (Appendix~\ref{sec:appa}). 

In the following, we present caloric curves and use them as a guide for the interpretation of how $T>0$ density profile evolve between MB limit and the fully degenerate state.
Another diagnostic tool for the same purpose is the Helmholtz free energy:
\begin{equation}\label{eq:87}
F=E-TS=U+W-TS,
\end{equation}
where $U$ is the kinetic energy, $E$ the internal energy, $S$ the entropy, and $W$ the (gravitational) potential energy.

\subsection{Kinetic energy and entropy}\label{sec:kinen-entro}
From the expressions developed earlier we infer the following integrals for the kinetic energy and the entropy:
\begin{equation}\label{eq:98}
\hat{U}\doteq\frac{U}{Nk_\mathrm{B}T_\mathrm{s}}=
\mathcal{D}_\sigma g_s\hat{T}^{\mathcal{D}/2+1}\frac{\mathcal{D}}{2}
\int_0^{\hat{R}} d\hat{r}\,\hat{r}^{\mathcal{D}_\sigma-1}
f_{\mathcal{D}/2+1}(\hat{z}),
\end{equation}
\begin{align}\label{eq:99}
\hat{S}\doteq\frac{S}{Nk_\mathrm{B}} &=
\mathcal{D}_\sigma g_s\hat{T}^{\mathcal{D}/2}
\int_0^{\hat{R}} d\hat{r}\,\hat{r}^{\mathcal{D}_\sigma-1} 
\nonumber \\
&\hspace{-5mm}\times\left[\left(\frac{\mathcal{D}}{2}+1\right)f_{\mathcal{D}/2+1}(\hat{z})
-\ln\hat{z}\,f_{\mathcal{D}/2}(\hat{z})\right],
\end{align}
implying
\begin{align}\label{eq:100}
\hat{U}-\hat{T}\hat{S} &=
-\mathcal{D}_\sigma g_s\hat{T}^{\mathcal{D}/2+1}
\int_0^{\hat{R}} d\hat{r}\,\hat{r}^{\mathcal{D}_\sigma-1}
 \nonumber \\
&\hspace{5mm}\times\Big[f_{\mathcal{D}/2+1}(\hat{z})-\ln\hat{z}\,f_{\mathcal{D}/2}(\hat{z})\Big].
\end{align}

\subsection{Potential energy}\label{sec:grav-self-ener}
The construction of the potential energy $W$ for self-gravitating clusters requires some thought.
In all cases $1 \leq \mathcal{D}_\sigma \leq \mathcal{D} =3$ we choose a reference state (pseudo-vacuum) different from the ground state (physical vacuum), namely the state with all particles confined to $0\leq r\leq r_\mathrm{c}$ at uniform density in real space. 
All differences $\Delta W$ between macrostates are independent of $r_\mathrm{c}$.

We calculate the differential $dW$ as work performed against gravity when a thin layer of mass is translocated from the reference density profile to the actual density profile, a method developed in the context of a self-gravitating lattice gas \cite{selgra}. 
This construction turns out to be also useful for BE clusters \cite{sgcbe}.
Alternative expressions for $W$, which are equivalent and derived from the viral theorem, can be found in Appendix~\ref{sec:appb}.

For FD clusters with planar symmetry $(\mathcal{D}_\sigma=1)$ we arrive at the expression,
\begin{align}\label{eq:109} 
& \hat{W} \doteq \frac{W}{Nk_\mathrm{B}T_\mathrm{s}}
=2\int_0^{\hat{R}} d\hat{r}_2\hat{r}_2\sigma_1(\hat{r}_2)\rho(\hat{r}_2)
-\frac{2}{3}\hat{r}_\mathrm{c}, \nonumber \\
& \sigma_1(\hat{r}_2) \doteq \int_0^{\hat{r}_2} d\hat{r}\rho(\hat{r})
=-\frac{\hat{T}}{2}\frac{\hat{z}'(\hat{r})}{\hat{z}(\hat{r})}.
\end{align}
The corresponding expression for clusters with cylindrical symmetry $(\mathcal{D}_\sigma=2)$ has a quite different look:
\begin{align}\label{eq:147}
& \hat{W} =4\int_0^{\hat{R}} d\hat{r}_2\,\hat{r}_2\,\sigma_2(\hat{r}_2)\,\rho(\hat{r}_2)\ln\left(\frac{\hat{r}_2}{\sqrt{\sigma_2(\hat{r}_2)}}\right)
 -\ln\hat{r}_\mathrm{c}, \nonumber \\
& \sigma_2(\hat{r}_2)\doteq 2\int_0^{\hat{r}_2}d\hat{r}\,\hat{r}\rho(\hat{r})
=-\frac{\hat{T}}{2}\,\hat{r}\,\frac{\hat{z}'(\hat{r})}{\hat{z}(\hat{r})}.
\end{align}
The logarithmic terms, characteristic for cylindrical symmetry, disappear for spherical symmetry $(\mathcal{D}_\sigma=3)$:
\begin{align}\label{eq:170} 
& \hat{W} 
=-6\int_0^{\hat{R}} d\hat{r}_2\hat{r}_2\sigma_3(\hat{r}_2)\rho(\hat{r}_2)
+\frac{6}{5}\hat{r}_\mathrm{c}^{-1}, \nonumber \\
& \sigma_3(\hat{r}_2)
=3\int_0^{\hat{r}_2} d\hat{r}\,\hat{r}^2\rho(\hat{r})
=-\frac{\hat{T}}{2}\,\hat{r}^2\,\frac{\hat{z}'(\hat{r})}{\hat{z}(\hat{r})}.
\end{align}
The dependence on $\mathcal{D}$ in all cases is contained in the profiles $\hat{z}(\hat{r})$ and $\rho(\hat{r})$.

\subsection{Caloric curves in $\mathcal{D}=3$}\label{sec:cal-D3}
An examination of caloric curves sets the stage for the analysis of $T>0$ density profiles in Sec.~\ref{sec:emerge}.
Our scaling convention suggests that we plot $\hat{\beta}\doteq
k_\mathrm{B}T_\mathrm{s}/k_\mathrm{B}T$ versus $\hat{E}\doteq
E/Nk_\mathrm{B}T_\mathrm{s}$.
The scaled radius of confinement $\hat{R}\doteq R/r_\mathrm{s}$ is a useful parameter.
A more detailed discussion of caloric curves can be found in Refs.~\cite{ptdimd,ijmpb}.
In Appendix ~\ref{sec:appa} we explain the different scaling conventions in use.
In the following, we highlight systems for $\mathcal{D}_\sigma=1,2,3$ in $\mathcal{D}=3$.
The symmetry of the cluster has a stronger impact on caloric curves than the dimensionality of the space.

\begin{figure}[htb]
\begin{center}
\includegraphics[width=.65\linewidth]{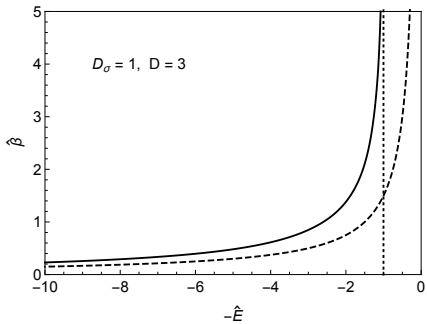} 
\end{center}
\caption{Caloric curves for the FD cluster (solid line) and MB cluster (dashed line) with planar symmetry and $\hat{R}=\infty$ in three-dimensional space. 
The dotted line marks the energy of the fully degenerate cluster 
identified in Sec.~\ref{sec:degen-planar-d3}. 
The MB result reflects the internal energy $E=\frac{5}{2}Nk_B T$ (see 
Appendix~\ref{sec:appb}).}
\label{fig:7}
\end{figure}

Caloric curves of clusters with planar symmetry are monotonically increasing across the complete range of inverse temperature (see Fig.~\ref{fig:7}). 
The range of negative internal energy has no lower limit, but reaches an upper limit, at $\hat{E}_\mathrm{min}=\frac{10}{9}\hat{r}_0$, where $\hat{r}_0=0.9015\ldots$ is the radius of the self-confined cluster identified in Sec.~\ref{sec:degen-planar-d3}.

The featureless structure of these caloric curves predicts that the cooling of FD clusters with planar symmetry incites a very gradual response throughout.
The cooling of MB clusters with planar symmetry is equally uneventful.
The FD caloric curves deviates from the MB caloric curve as the exclusion principle comes into play.
Effects of wall confinement (not shown) are significant only when the space is
very tight or 
the temperature very high (for confined systems, $E\sim
\frac{\mathcal{D}}{2}Nk_B T$ when $T\rightarrow +\infty$). 

\begin{figure}[htb]
\begin{center}
\includegraphics[width=.48\linewidth]{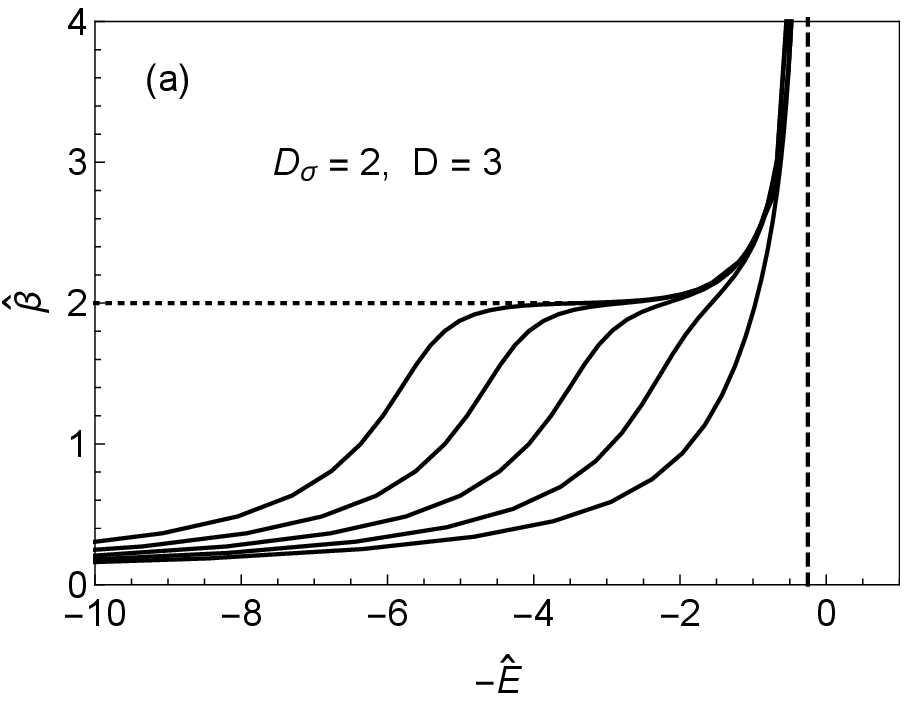}\hspace{2mm}%
\includegraphics[width=.48\linewidth]{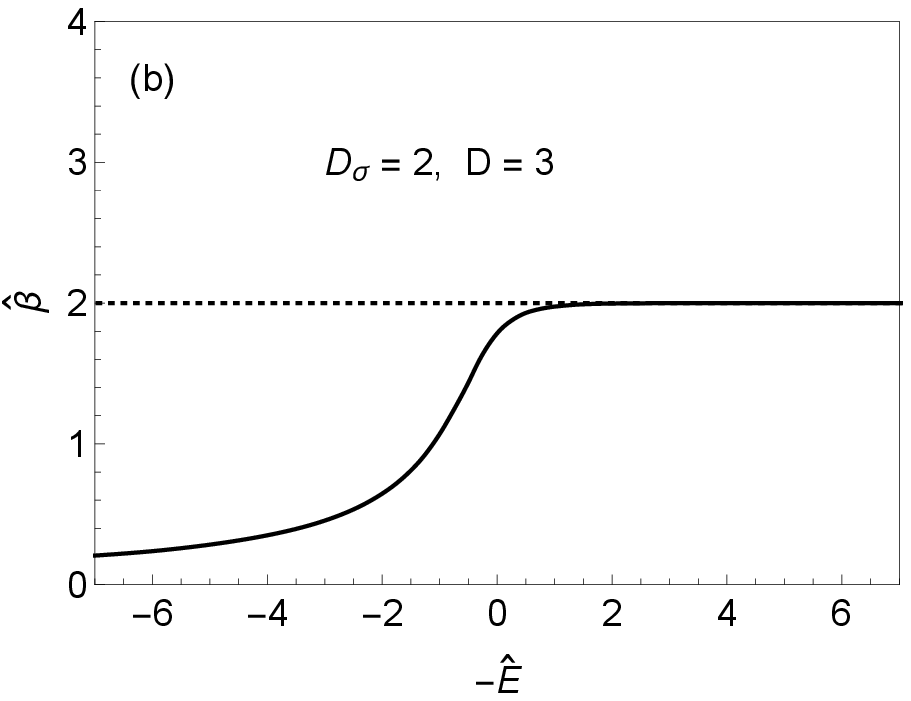}
\end{center}
\caption{ (a) Caloric curves for FD clusters with confining radius
$\hat{R}=1.50, 4.70, 14.9, 47.0, 149$ (right to left) and $\hat{R}=\infty$ (dotted line).
The dashed lines marks the energy of the fully degenerate FD cluster identified in Sec.~\ref{sec:degen-cylind-d3}.
(b) Caloric curve for MB clusters of any $\hat{R}$.
The dotted line marks temperature $\hat{T}_\mathrm{MB}=\frac{1}{2}$ at which the MB clusters collapse.}
\label{fig:8}
\end{figure}

The caloric curves for FD clusters with cylindrical symmetry are still strictly monotonic as shown in Fig.~\ref{fig:8}(a), but they are no longer featureless. 
There is a plateau in the vicinity of the characteristic temperature $\hat{T}_\mathrm{MB}=\frac{1}{2}$, where the MB cluster is known to undergo a gravitational collapse as evident in Fig.~\ref{fig:8}(b).
Here the plateau reaches all the way to $\hat{E}\to-\infty$, where the MB cluster has contracted to a point.

The FD caloric curves with the plateau feature represent a phenomenon of incipient gravitational collapse, more or less gently arrested by the counteracting exclusion pressure.
However, there is no hint of gravitational collapse if the average inter-particle distance is comparable to thermal wavelength already at temperatures near $\hat{T}_\mathrm{MB}$.
This is the case under tight confinement.
In summary, despite the added structure, the strict monotonicity of all FD caloric curves for clusters with cylindrical symmetry rules out any mechanical instabilities or phase transitions.

\begin{figure}[htb]
  \begin{center}
 \includegraphics[width=.48\linewidth]{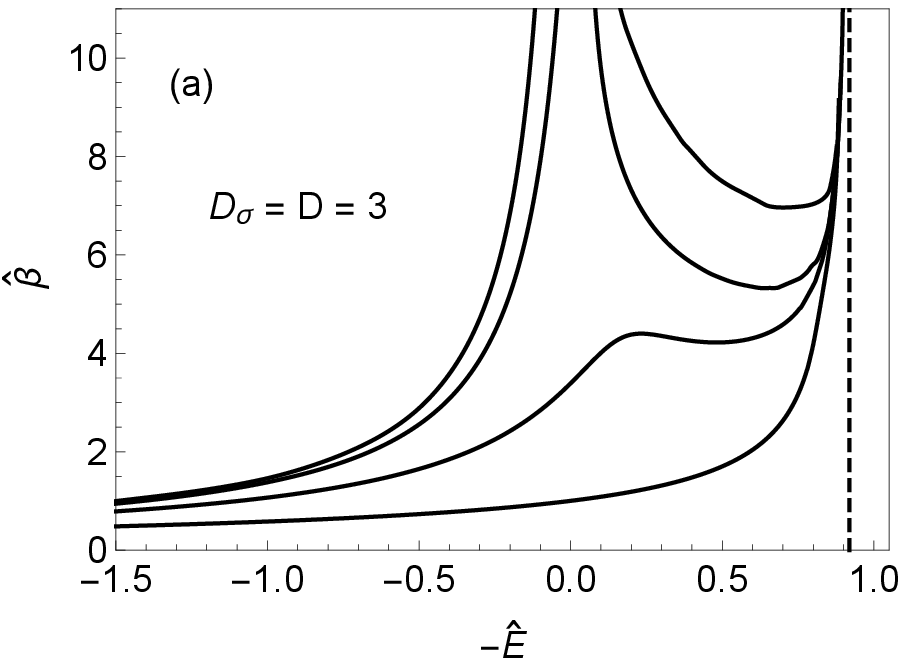}  
 \includegraphics[width=.48\linewidth]{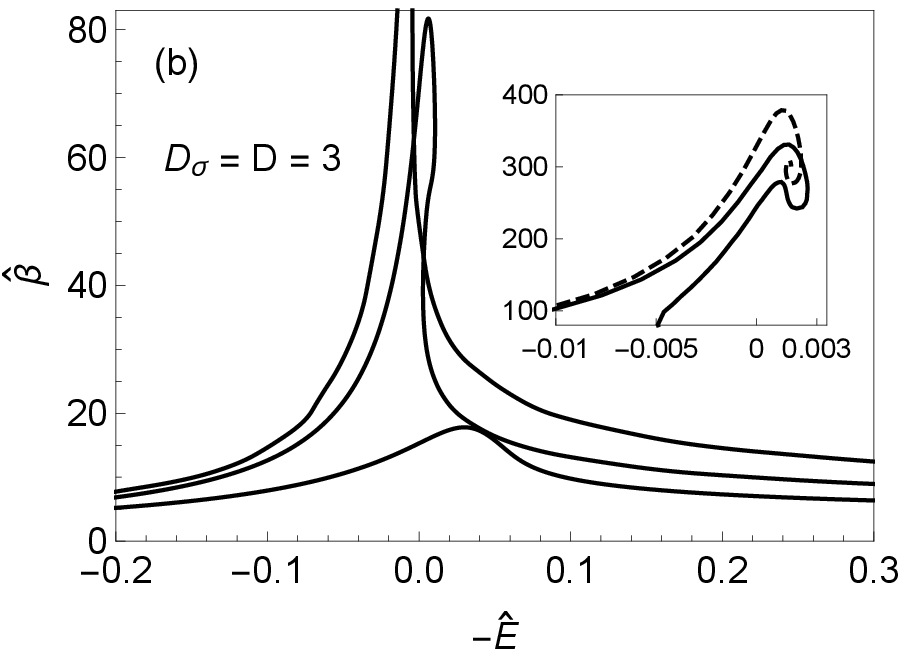}  
\end{center}
\caption{Caloric curves of FD clusters with spherical symmetry with confining radius (a) $\hat{R}=0.648, 3.01,14.0, 64.8$, (b) $\hat{R}=14.0, 64.8, 301$.
The inset zooms into a part of the curve for $\hat{R}=301$  (named dinosaur's neck \cite{ijmpb}) not shown in the main plot and compares it with a similar feature of the MB caloric curve (dashed line).}
  \label{fig:9}
\end{figure}

A whole new level of drama is on display in the caloric curves for FD clusters 
with spherical symmetry (Fig.~\ref{fig:9}).
Here we see everything that we already identified and much more.
In very tight quarters $(\hat{R}\ll1)$, the caloric curve is monotonic and featureless as is the case universally in planar clusters.
As we relax the confinement by increasing $\hat{R}$, a shoulder makes its appearance, which is reminiscent of caloric curves of cylindrical clusters.

Upon further widening of the space to which the FD gas is confined, hitherto unseen structures emerge.
The first novel structural feature that makes its appearance is local maximum followed by a local minimum, two points with zero slope.
They signal the presence of multiple macrostates with different internal energy at the same temperature. 
The points of zero slope indicate locations of mechanical instability in the canonical ensemble.

A further loosening of the confinement introduces points on the caloric curve with infinite slope, indicating the presence of multiple macrostates  at different temperature with the same internal energy.
Such points are well known to be associated with mechanical instabilities in the
microcanonical ensemble \cite{ijmpb}.
Both types of instabilities point to hysteretic behavior of one-phase macrostates (Sec.~\ref{sec:emerge}), but they also point to the possibility of phase coexistence (Sec.~\ref{sec:phase coex}).
 
%
\section{Emergent degeneracy}\label{sec:emerge}
%
We have seen that density profiles of fully
degenerate finite-mass FD clusters have compact support. 
This is no longer the case for equilibrium macrostates at nonzero temperature.
The cluster surfaces becomes fuzzy -- an attribute related to the absence of short-range cohesive forces in our modeling.
The density profile acquires a tail out to infinite distances from the center of the cluster.
Wall confinement becomes necessary to equilibrate finite-mass clusters at nonzero temperature in some cases.

\subsection{Asymptotics of self-confined clusters}\label{sec:asymp-self-conf}
It is instructive to take a look at the asymptotic decay laws of density profiles up front. 
The exact analysis, which follows the approach of Ref.~\cite{selgra}, starts from
\begin{equation}\label{eq:400} 
\frac{\hat{z}'(\hat{r})}{\hat{z}(\hat{r})}\,\hat{r}^{\mathcal{D}_\sigma-1}
=-\frac{2\mathcal{D}_\sigma}{\hat{T}}\int_0^{\hat{r}}d\hat{r}'
\hat{r}'^{\mathcal{D}_\sigma-1}\rho(\hat{r}'),
\end{equation}
inferred from (\ref{eq:6}), (\ref{eq:8}), and (\ref{eq:1a}).
With (\ref{eq:24})  we can write,
\begin{equation}\label{eq:401} 
\frac{\hat{z}'(\hat{r})}{\hat{z}(\hat{r})}\,\hat{r}^{\mathcal{D}_\sigma-1}
=-\frac{2}{\hat{T}}\left[1-\int_{\hat{r}}^\infty d\hat{r}'
\hat{r}'^{\mathcal{D}_\sigma-1}\rho(\hat{r}')\right].
\end{equation}
The finite-mass condition for density profiles with power-law decay is a lower bounday for the exponent:
\begin{equation}\label{eq:402}
\rho(\hat{r})\sim\hat{r}^{-\eta},\quad \eta>\mathcal{D}_\sigma.
\end{equation}
This condition ensures that the integral in (\ref{eq:401}) becomes negligibly small at large $\hat{r}$.
In the low-density asymptotic regime we have $\hat{z}'/\hat{z}=\rho'/\rho$.
We thus extract from (\ref{eq:401}) the limit,
\begin{equation}\label{eq:403}
 \lim_{\hat{r}\to\infty}\frac{\rho'(\hat{r})}{\rho(\hat{r})}\,
 \hat{r}^{\mathcal{D}_\sigma-1}=-\frac{2}{\hat{T}},
\end{equation}
which only depends on the symmetry of the cluster, but not on the dimensionality of the space.
The solution of (\ref{eq:403}) yields exponential asymptotics for planar clusters,
\begin{equation}\label{eq:404}
\rho(\hat{r})_\mathrm{as} \sim e^{-2\hat{r}/\hat{T}}\quad 
:~ \mathcal{D}_\sigma=1,
\end{equation}
and power-law asymptotics for cylindrical clusters,
\begin{equation}\label{eq:405}
\rho(\hat{r})_\mathrm{as} \sim \hat{r}^{-2/\hat{T}}\quad 
:~ \mathcal{D}_\sigma=2.
\end{equation}

The finite-mass condition (\ref{eq:402}) restricts the temperature range for the power-law asymptotics (\ref{eq:405}) to $0<\hat{T}<1$.
We shall see that the self-confinement condition for cylindrical clusters is more stringent and restricts power-law asymptotics to $0<\hat{T}<\frac{1}{2}$.

Self-confined clusters with spherical symmetry at nonzero temperature only exist if they have infinite mass.
The leading asymptotic decay of the density profile is independent of
temperature in this case \cite{chandrabook}:
\begin{equation}\label{eq:411}
\rho(\hat{r})_\mathrm{as} \sim 2\hat{r}^{-2}\quad 
:~ \mathcal{D}_\sigma=3.
\end{equation}

\subsection{Planar symmetry}\label{sec:plan-sym}
It is well known that an unconfined MB cluster with planar symmetry remains
stable against evaporation or gravitational collapse at any nonzero temperature.
The exact density profile
is \cite{spitzer,camm,rybicki,kl,sc,cmct,selgra}
\begin{equation}\label{eq:210} 
\rho(\hat{r})_\mathrm{MB}=\frac{1}{\hat{T}}\,\mathrm{sech}^2\!\left(\frac{\hat{r}}{\hat{T}}\right).
\end{equation}
The exponential FD asymptotics (\ref{eq:404}) is also realized in the MB profile (\ref{eq:210}), as expected.
With decreasing $\hat{T}$, $\rho(\hat{r})_\mathrm{MB}$ gradually becomes narrower and more strongly peaked at the central plane of the cluster, approaching a $\delta$-function in the limit $\hat{T}\to0$.
Deviations of the FD density profile from the MB result (\ref{eq:210}) are expected to emerge gradually.
The exclusion principle kicks into action when the (local) average inter-particle distance becomes comparable to the thermal wavelength.
This criterion is first met near the center of the cluster, where the density is highest.

\begin{figure}[htb]
  \begin{center}
 \includegraphics[width=.48\linewidth]{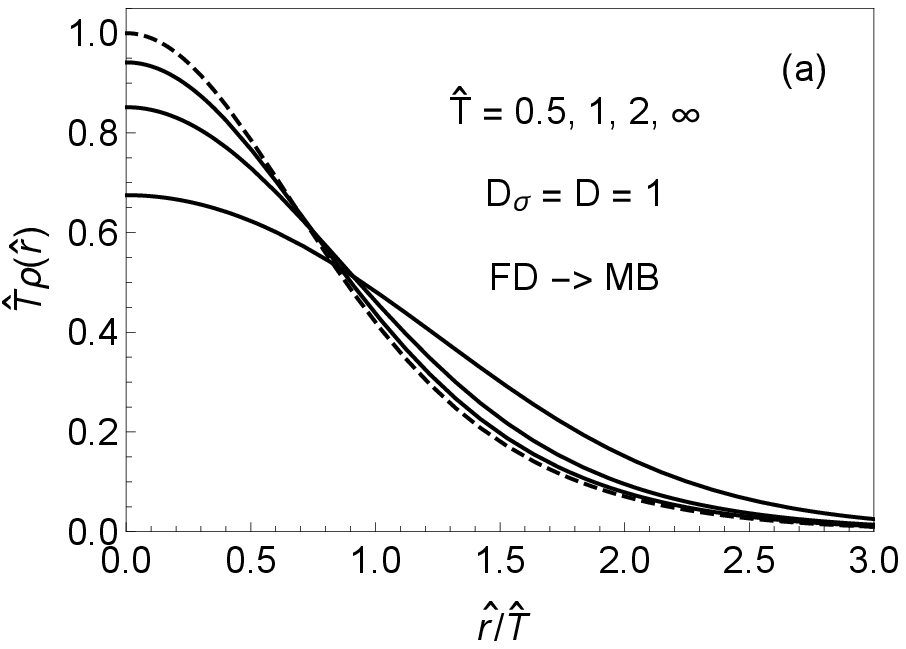}  
 \includegraphics[width=.48\linewidth]{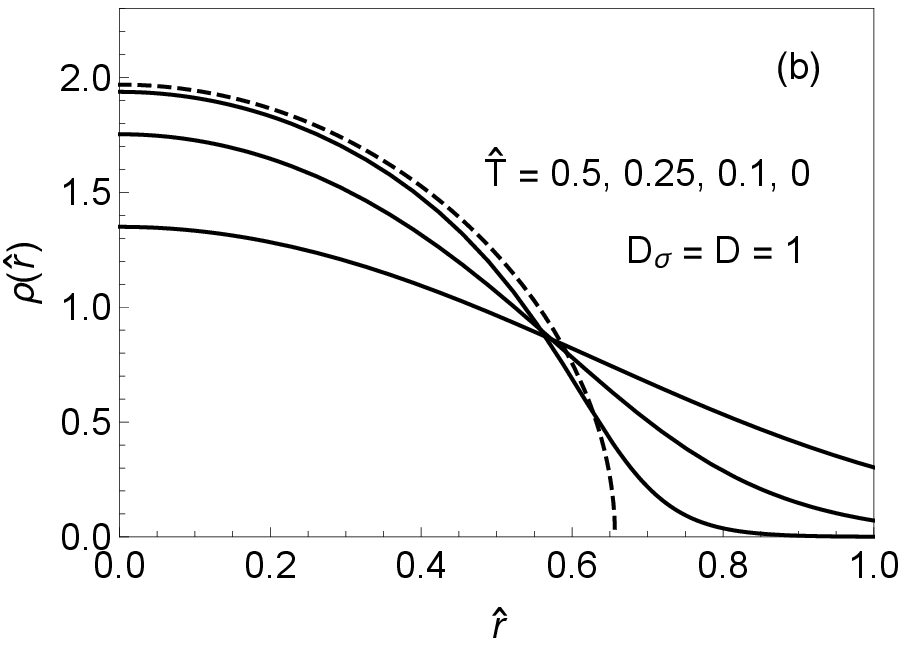}  
\end{center}
\caption{(a) Rescaled density profiles of the FD gas in $\mathcal{D}_\sigma=\mathcal{D}=1$ at high $\hat{T}$. The dashed line represents the MB profile (\ref{eq:210}), which is independent of $\hat{T}$ in rescaled units.
(b) Density profiles of FD clusters at  low $\hat{T}$.
The dashed line represents the $\hat{T}=0$ profile from Sec.~\ref{sec:degen-planar-d1}.}
  \label{fig:10}
\end{figure}

The central density of the FD cluster is indeed suppressed relative to that of an MB cluster as shown in 
Fig.~\ref{fig:10}(a).
The dashed line represents the universal MB density profile in rescaled units.
The solid lines illustrate how the FD density profile (for $\mathcal{D}=1$) deviates from it as the temperature is being lowered from a high value. The MB result is independent of $\mathcal{D}$.
The FD deviations are similar in $\mathcal{D}=2,3$ (not shown), albeit somewhat slower with increasing $\mathcal{D}$, as might be expected.

\begin{figure}[htb]
  \begin{center}
 \includegraphics[width=.48\linewidth]{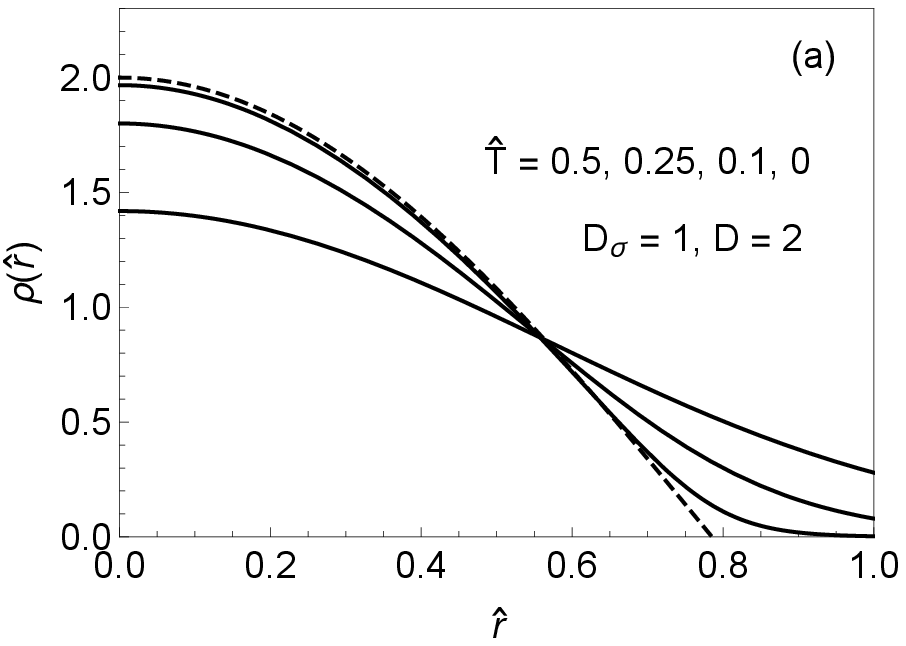}  
 \includegraphics[width=.48\linewidth]{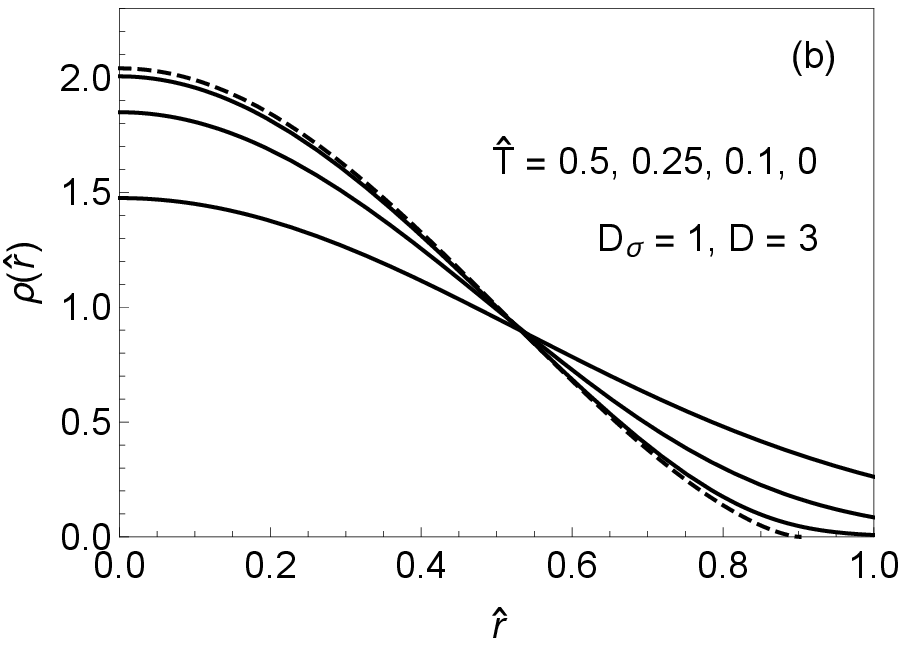}  
\end{center}
\caption{Density profiles of planar FD gas in (a) $\mathcal{D}=2$ and (b) $\mathcal{D}=3$ at low $\hat{T}$. The dashed lines represent the $\hat{T}=0$ profiles from Sec.~\ref{sec:degen-planar-d2} and Sec.~\ref{sec:degen-planar-d3}, respectively.}
  \label{fig:11}
\end{figure}

The dependence on space dimension $\mathcal{D}$ of the FD profiles becomes more conspicuous at low $\mathcal{D}$ as shown in Figs.~\ref{fig:10}(b) and \ref{fig:11}. 
The solutions of the ODE (\ref{eq:23}) for the fugacity at $\hat{T}>0$ converge neatly toward the solutions of the ODE (\ref{eq:42}) for the chemical potential at $\hat{T}=0$.
The density of particles at large distances is exponentially suppressed according to (\ref{eq:404}) in all $\mathcal{D}$, but the limiting $\hat{T}=0$ profile strongly depneds on $\mathcal{D}$.

\subsection{Cylindrical symmetry}\label{sec:cylin-sym}
MB gas clusters with cylindrical symmetry, which we again use as a benchmark, are vulnerable to both particle escape and gravitational collapse.
In a wall-confined space of radius $\hat{R}$ -- a disk 
in $\mathcal{D}=2$ or a cylinder in $\mathcal{D}=3$ -- MB clusters are stable
against gravitational collapse for temperatures above the threshold value
\cite{cf,stodolkiewicz,ostriker,salzberg,klb,paddy2d,aly,ar,ap,
sc,cmct,bppv,virialD,selgra},
\begin{equation}\label{eq:218} 
\hat{T}_\mathrm{MB}=\frac{1}{2}.
\end{equation}
The exact density profile reads \cite{sc,selgra}:
\begin{equation}\label{eq:219} 
 \rho(\hat{r})_\mathrm{MB}=\frac{1}{\hat{R}^2}\frac{2\hat{T}\big(2\hat{T}-1\big)}
 {\big[(\hat{r}/\hat{R})^2+2\hat{T}-1\big]^2}.
\end{equation}
A one-parameter family of density profiles exists in the combined limit,
\begin{equation}\label{eq:406} 
 \hat{T}\to\hat{T}_\mathrm{MB},\quad \hat{R}\to\infty,\quad 
 \frac{\hat{T}^2}{2\hat{R}^2(2\hat{T}-1)}\to c>0,
\end{equation}
at the border between collapse or escape \cite{cmct,selgra}:
\begin{equation}\label{eq:407} 
 \rho(\hat{r})_\mathrm{MB}=\frac{4c}{\hat{T}_\mathrm{MB}}
 \left[1+2c\left(\frac{\hat{r}}{\hat{T}_\mathrm{MB}}\right)^2\right]^{-2}.
\end{equation}
This includes the collapsed state $(c\to\infty)$ and the evaporated state $(c\to0)$.
The density profile (\ref{eq:407}) of this precarious MB state without wall confinement does exhibit the power-law asymptotics (\ref{eq:405}) for the FD cluster, albeit only at the threshold temperature.
FD clusters with cylindrical symmetry are equally vulnerable to particle escape, but not to gravitational collapse.
Whereas the MB profile (\ref{eq:219}) acquires an unlimited central density as $\hat{T}$ approaches $\hat{T}_\mathrm{MB}$ from above, the FD particles resist such squeezing on account of the exclusion principle.

\begin{figure}[htb]
  \begin{center}
 \includegraphics[width=.48\linewidth]{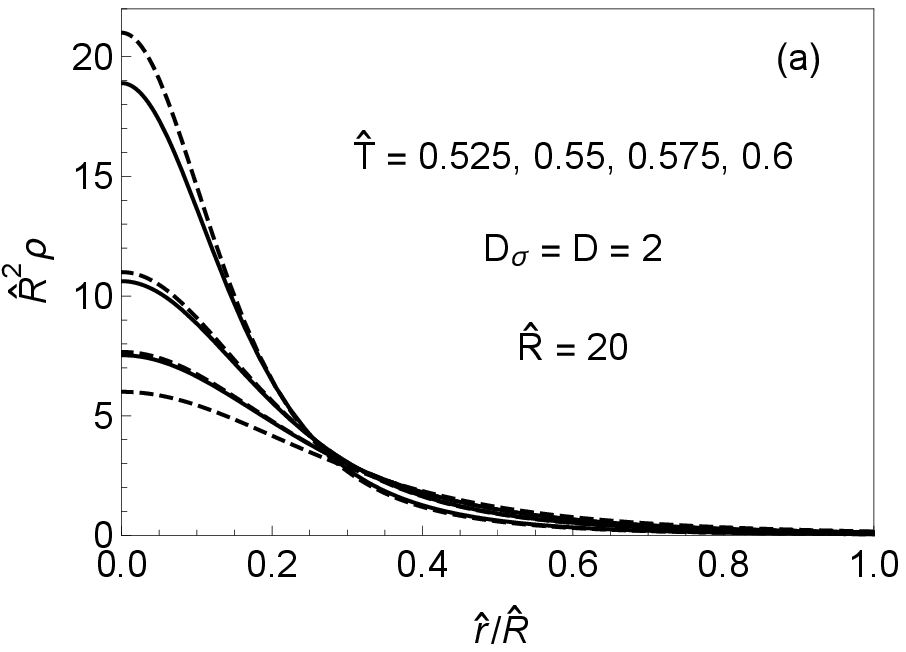}  
 \includegraphics[width=.48\linewidth]{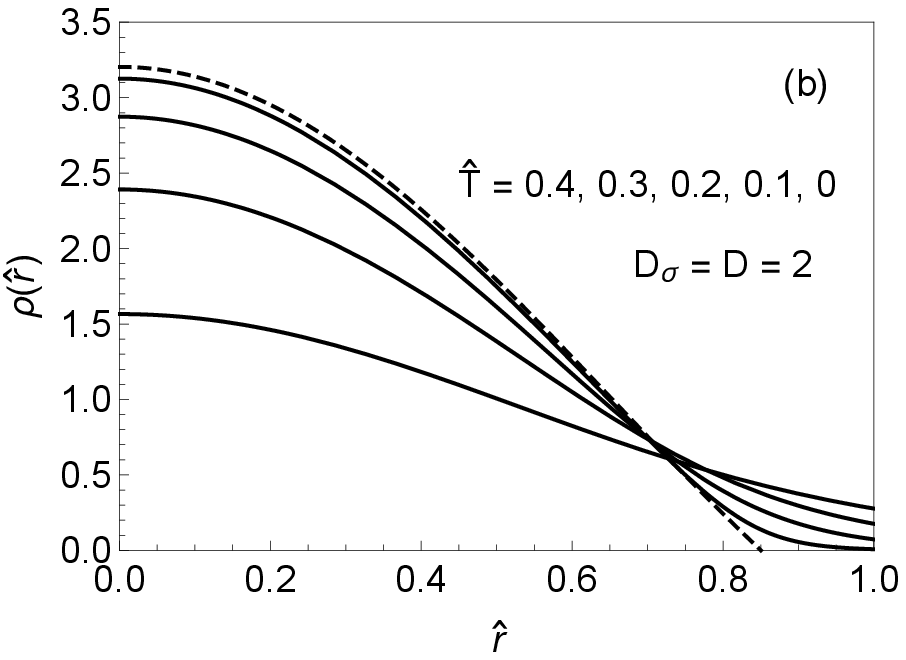}  
\end{center}
\caption{(a) Rescaled FD density profiles in $\mathcal{D}_\sigma=\mathcal{D}=2$ at $\hat{T}>\hat{T}_\mathrm{MB}$ in a disk-shaped space. The dashed line represents the MB profile (\ref{eq:219}). 
(b) Scaled FD density profiles at $\hat{T}<\hat{T}_\mathrm{MB}$ in an unconfined space.
The dashed line represents the $\hat{T}=0$ profile from Sec.~\ref{sec:degen-cylind-d2}.}
  \label{fig:12}
\end{figure}

For the sake of brevity, we focus on dimension $\mathcal{D}=2$.
The results for $\mathcal{D}=3$ are very similar except in the limit $\hat{T}\to0$ (Sec.~\ref{sec:degen-cylind-d2}).
Connecting the FD solutions of (\ref{eq:23}) with the MB result (\ref{eq:219}) graphically again requires some rescaling, as shown in Fig.~\ref{fig:12}(a).
The choice of a large $\hat{R}$ facilitates a convergence between FD and MB profiles not far above $\hat{T}_\mathrm{MB}$.
The approach of the FD profiles $T>0$ toward the $\hat{T}=0$ profile from Sec.~\ref{sec:degen-cylind-d2} is shown in Fig.~\ref{fig:12}(b) using different scales.

\begin{figure}[htb]
  \begin{center}
 \includegraphics[width=.48\linewidth]{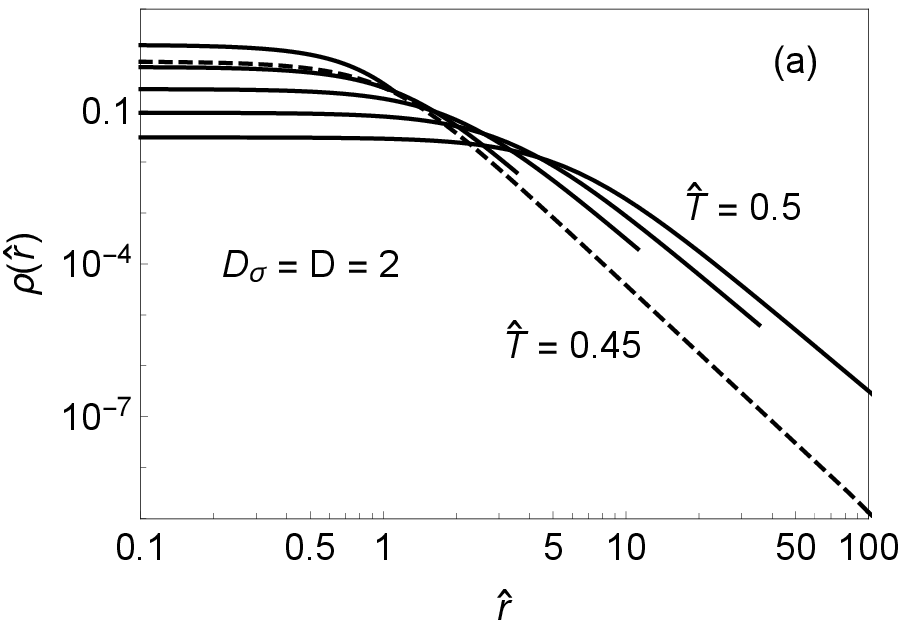}  
 \includegraphics[width=.48\linewidth]{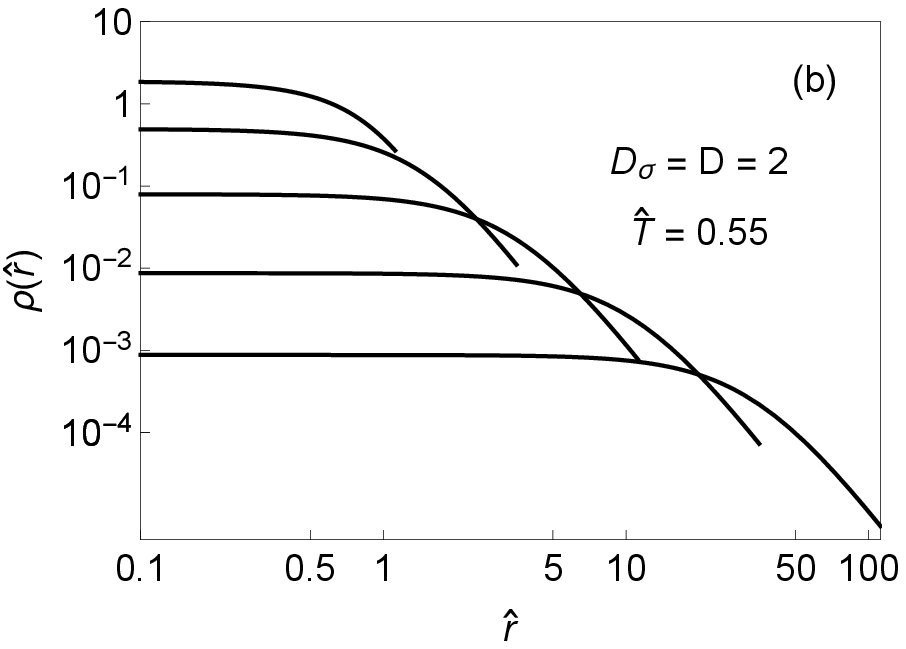}  
\end{center}
\caption{Density profiles for the FD cluster with $\mathcal{D}_\sigma=\mathcal{D}=2$ at (a) $\hat{T}=0.45$ (dashed line), $\hat{T}=0.5$ (solid lines) and (b) $\hat{T}=0.55$. The radius of wall confinement, $\hat{R}=1.12, 3.54, 11.2, 35.4, 112$ is indicated by the endpoint of each solid curve.}
  \label{fig:13}
\end{figure}

The presence of self-confinement at $\hat{T}<\hat{T}_\mathrm{MB}$ and its absence at $\hat{T}>\hat{T}_\mathrm{MB}$ are illustrated in Fig.~\ref{fig:13} from a different angle \cite{selgra}.
When we widen the space for the gas at constant subcritical temperature $\hat{T}=0.45$ by isothermally increasing the radius $\hat{R}$, the profile shows virtually no response [dashed line in Fig.~\ref{fig:13}(a)]. 
At that temperature, self-confinement is robust and the observed power-law decay is in accord with the asymptotics (\ref{eq:405}).

When the wall confinement is quasi-statically and isothermally relaxed at criticaity, the cluster responds differently as shown by the solid lines in Fig.~\ref{fig:13}(a).
A power-law tail,  $\sim\hat{r}^{-4}$, survives and becomes increasingly conspicuous for large $\hat{R}$, in agreement with the asymptotics (\ref{eq:405}).
What makes the critical profle qualitatively different from the subcritical profiles is the presence of a flat portion of increasing width and decreasing height around the center of the cluster.
Self-confinement is no longer operational.

Whereas self-confinement is already absent at criticality, the structure of density profiles is yet different at higher temperatures such as shown in Fig.~\ref{fig:13}(b) for $\hat{T}=0.55$.
Here the flat portion becomes more dominant, extending over a wider region, and the power-law tail has disappeared.
The only evidence of gravity is the reduced density near the wall.

There are no abrupt changes when a wall-confined FD cluster with cylindrical symmetry is heated up or cooled down quasi-statically.
This trait is shared with planar clusters, where wall-confinement is not even necessary.
An unconfined FD cluster with cylindrical symmetry will spread out gradually when heated up. 
Its density will become low everywhere before $\hat{T}_\mathrm{MB}$ is reached,
at which point it will behave MB like.

Confinement with a wide radius $\hat{R}$ makes it possible to observe a crossover between density profiles that can be interpreted as incipient escape.
Within a relatively short interval of rising $\hat{T}$, the density profile changes from a core/halo variety to flat variety (Fig.~\ref{fig:13}).
This crossover has also been seen in the caloric curves (Fig.~\ref{fig:8}).
We shall see next that a switch from cylindrical to spherical symmetry impacts both the MB and the FD clusters qualitatively, but in quite different ways.

\subsection{Spherical symmetry}\label{sec:sphe-sym}
Here the inequivalence of ensembles matters, as the structure of caloric curves made clear (Sec.~\ref{sec:caloric}).
For the sake of brevity, our focus will be on the canonical ensemble. 
Analogous reasoning produces corresponding results for the microcanonical ensemble.
Spherical gas clusters of finite mass at nonzero temperature must be stabilized against escape by wall confinement.
This attribute is independent of statistics.
For the stability against gravitational collapse, the statistics does matter, of course.
The finite-mass MB cluster is only stable above a threshold temperature which depends  on the radius of confinement \cite{emden,lbw,aaiso},
\begin{equation}\label{eq:TC} 
\hat{T}_\mathrm{C}=\frac{\tilde{T}_\mathrm{C}}{\hat{R}},\quad 
\tilde{T}_\mathrm{C}=0.794422\ldots.
\end{equation}
Unlike in the case of cylindrical symmetry (Sec.~\ref{sec:cylin-sym}), where, on the verge of collapse, the MB gas is highly concentrated at the center of the cluster, the density profile of a spherical MB cluster at the point of collapse is still broad with the gas pushing against the wall.

In the caloric curves of cylindrical FD clusters we have noted an incipient instability for large $\hat{R}$ around $\hat{T}_\mathrm{MB}$, where the MB cluster undergoes a real instability.
We shall see that spherical FD clusters under tight confinement behave similarly, but now in the vicinity of $\hat{T}_\mathrm{C}$, where the spherical MB cluster collapses.
Under more loose confinement, by contrast, the spherical FD cluster exhibits a real instability, again near $\hat{T}_\mathrm{C}$, yet very unlike the MB instability.

\begin{figure}[htb]
  \begin{center}
 \includegraphics[width=.48\linewidth]{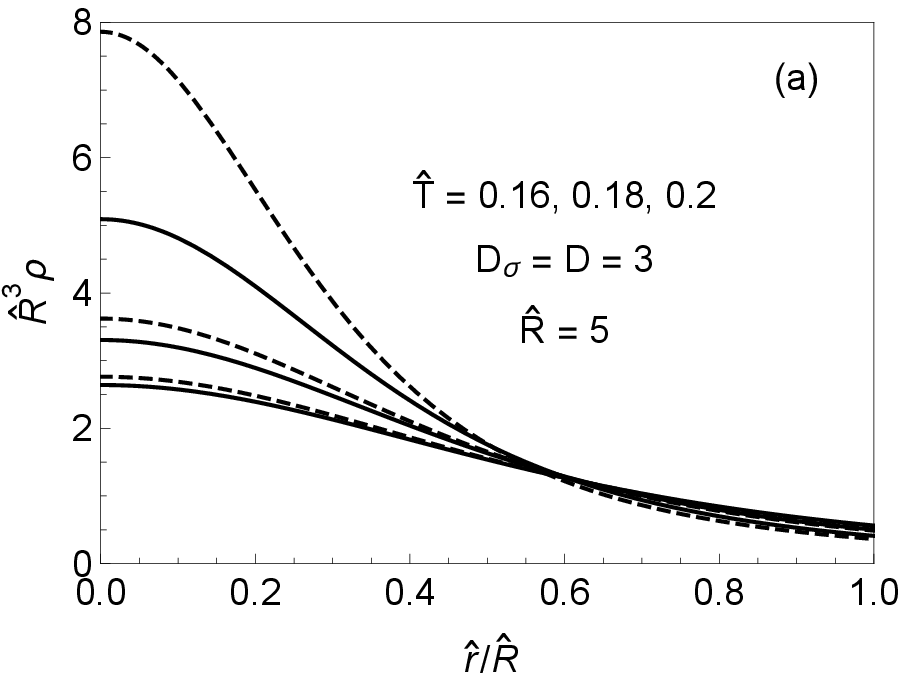}  
 \includegraphics[width=.48\linewidth]{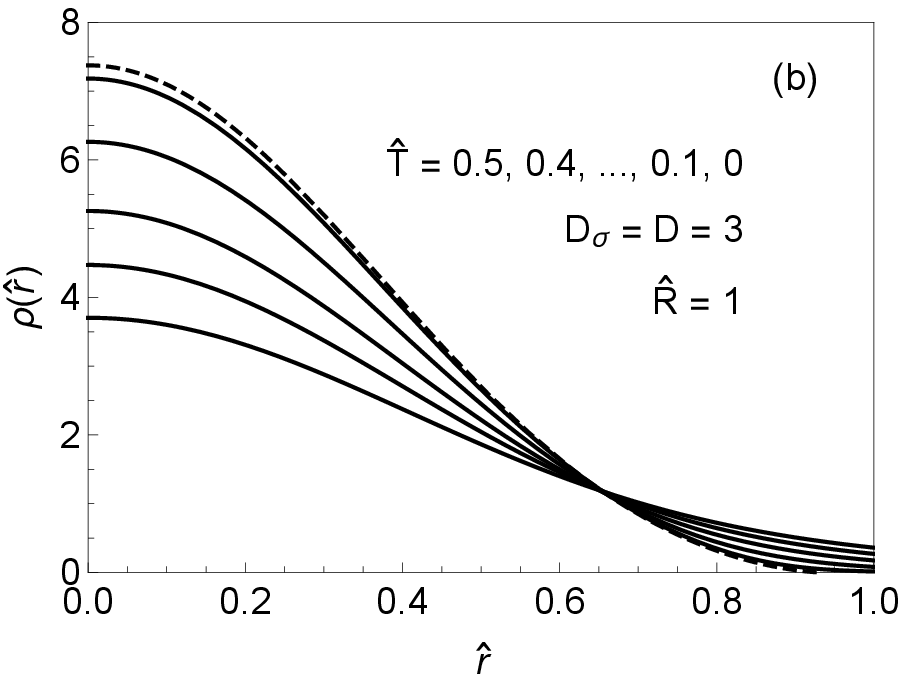}  
\end{center}
\caption{ (a) Density profiles of a confined FD gas (solid curves) and MB gas (dashed curves) in $\mathcal{D}_\sigma=\mathcal{D}=3$ at temperatures near $\hat{T}_\mathrm{C}$. (b) Density profile of an FD gas on approach to the fully degenerate limiting case (dashed curve). Wall confinement turns into self confinement at $\hat{T}=0$ .}
  \label{fig:14}
\end{figure}

We begin our analysis of the spherical FD cluster with finite mass by establishing the contacts with the high-$T$ and low-$T$ anchor points. 
In Fig.~\ref{fig:14}(a) we compare density FD and MB profiles as $\hat{T}$ approaches $\hat{T}_\mathrm{C}$ from above. 
The defenses against compression are getting weaker in the MB gas and stronger in the FD gas.
At the lowest $\hat{T}$, the collapse of the former is imminent.
In Fig.~\ref{fig:14}(b) we show the convergence of the latter to the fully degenerate profile discussed in Sec.~\ref{sec:degen-sphere-d3}.

For the transformation of high-temperature profiles [Fig.~\ref{fig:14}(a)] into
low-temperature profiles [Fig.~\ref{fig:14}(b)] and vice versa we distinguish
between a regime (i) of tight confinement 
and a regime (ii) of loose confinement.
In regime (i) for small $\hat{R}$, the density profile is unique at all temperatures and evolves gradually between the high-$\hat{T}$ MB-like  profile and the self-confined Fermi ball in the low-$\hat{T}$ limit.
In regime (ii) for large $\hat{R}$, on the other hand, there exists a temperature interval with multiple coexisting solutions of (\ref{eq:23}).
No smooth and continuously varying density profile across that interval exists.
Singular behavior is inevitable when the cluster is quasi-statically heated up or cooled down.

\begin{figure}[t]
  \begin{center}
 \includegraphics[width=.48\linewidth]{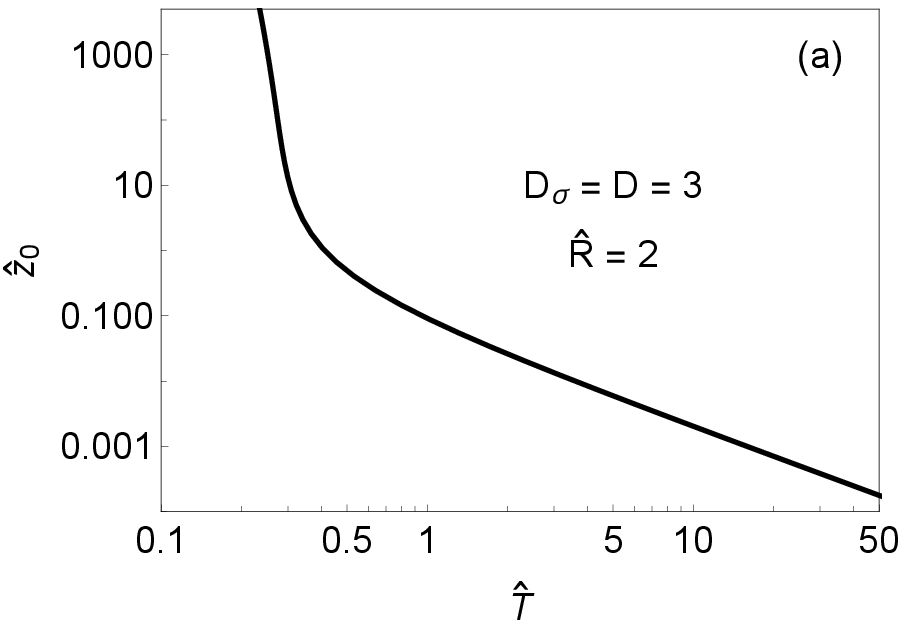}  
 \includegraphics[width=.48\linewidth]{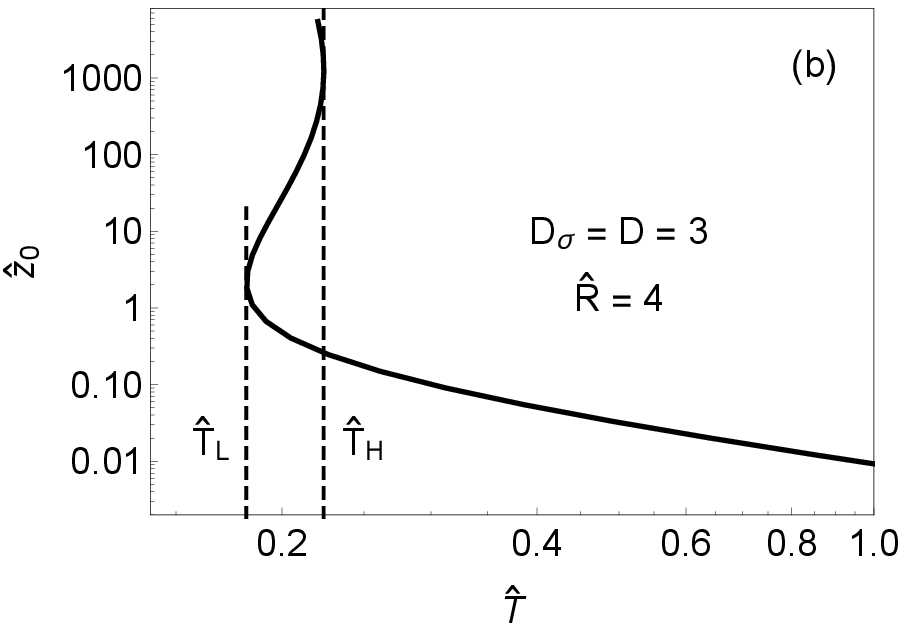}  
\end{center}
\caption{Fugacity $\hat{z}_0$ at the center of the cluster versus scaled 
temperature $\hat{T}$ for the cases with $\hat{R}=2$ and $\hat{R}=4$. The values
of local extrema in (b) are $\hat{T}_\mathrm{L}\simeq0.182$ and
$\hat{T}_\mathrm{H}\simeq0.224$. Both curves continue with negative slope toward $\hat{T}=0$ and $\hat{z}_0\to\infty$.}
  \label{fig:15}
\end{figure}

We have located the border between the two regimes at $\hat{R}_\mathrm{c}\simeq2.655$.
In the following, we compare one case from each regime.
We pick $\hat{R}=2$ for regime (i) and $\hat{R}=4$ for regime (ii).
In case (i), $\hat{z}_0$ is a single-valued monotonic function of $\hat{T}$ and in case (ii) a multiple-valued monotonic function (Fig.~\ref{fig:15}).
This means that the density profile is unique at all temperature in case (i) and all temperatures $\hat{T}<\hat{T}_\mathrm{L}$ or $\hat{T}>\hat{T}_\mathrm{H}$ in case (ii).
However, between the temperatures $\hat{T}_\mathrm{L}$ and $\hat{T}_\mathrm{H}$ in case (ii), there exist three coexisting profiles. 

\begin{figure}[htb]
  \begin{center}
 \includegraphics[width=.48\linewidth]{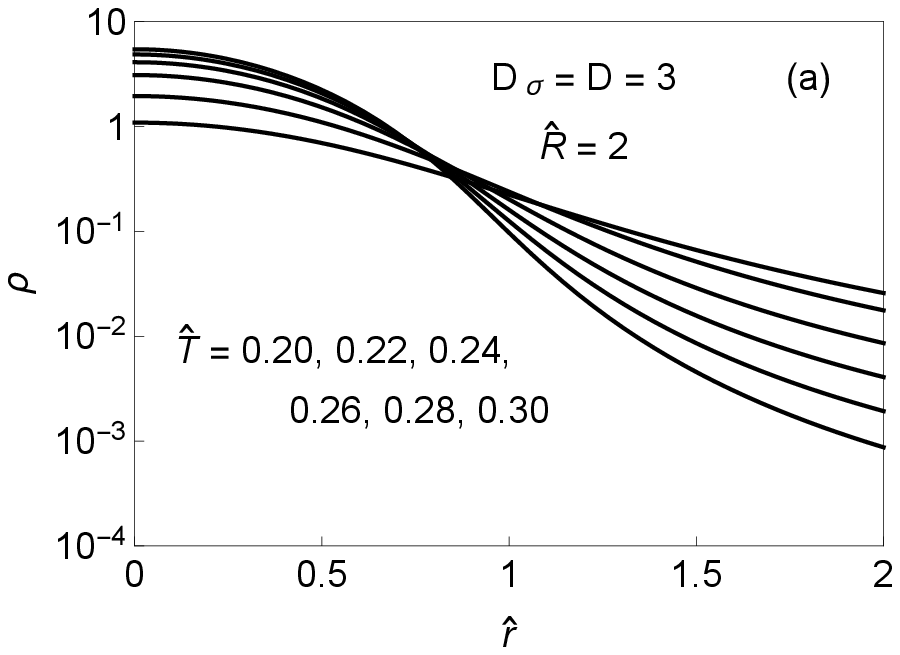}  
 \includegraphics[width=.48\linewidth]{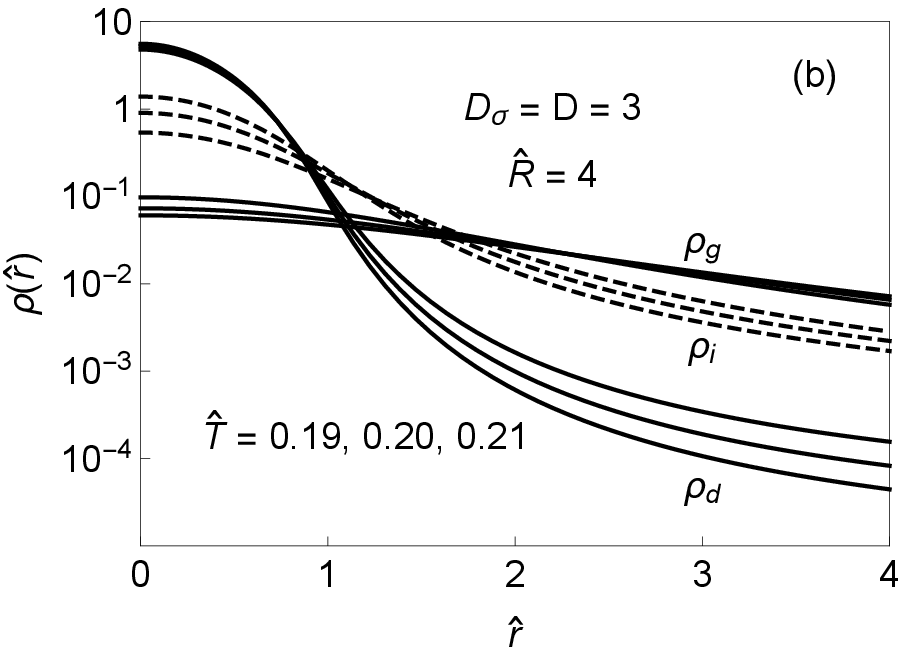}  
\end{center}
\caption{Density profiles at specific scaled temperatures $\hat{T}$ for (a) $\hat{R}=2$ and (b) $\hat{R}=4$. The three sets of profiles in (b)  named $\rho_\mathrm{g}$, $\rho_\mathrm{i}$, and $\rho_\mathrm{d}$, represent gaseous profiles, (unstable) intermediate profiles, and profiles with a degenerate core and a gaseous halo, respectively. The intercept of the solid (dashed) curves decreases (increases) with rising $\hat{T}$.}
  \label{fig:16}
\end{figure}

In Fig.~\ref{fig:16} we show how the density profiles evolve across a range of temperatures for cases (i) and (ii).
Panel (a) depicts profiles across an interval of $\hat{T}$ where the most rapid (yet still gradual) change occurs.
As $\hat{T}$ is being lowered, dominance shifts gradually from thermal fluctuations (dispersing agent) to gravity (aggregating agent).
The latter is, in turn, counteracted by the exclusion principle (agent akin to
steric repulsion).

In panel (b) we show three coexisting density profiles for
temperatures from the interval $\hat{T}_\mathrm{L}<\hat{T}<\hat{T}_\mathrm{H}$.
In this case, lowering $\hat{T}$ has a more dramatic effect.
At $\hat{T}>\hat{T}_\mathrm{H}$ the only solution is a flat profile
$\rho_\mathrm{g}$, similar to the ones shown.
The profile $\rho_\mathrm{g}$ represents a gaseous phase.
At $\hat{T}_\mathrm{H}$ two additional solutions emerge.
Initially they are identical, then evolve differently.
The solution $\rho_\mathrm{d}$ represents a degenerate cluster surrounded by a
gaseous halo with a fuzzy interface.
The solution $\rho_\mathrm{i}$ (shown dashed) represents an unstable
intermediate profile.

\begin{figure}[t]
\begin{center}
\includegraphics[width=.65\linewidth]{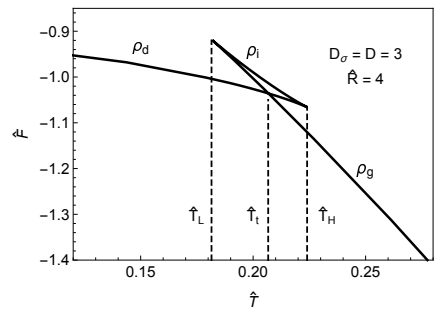} 
\end{center}
\caption{Free energy versus temperature of the macrostates with central fugacities as shown in Fig.~\ref{fig:15}(b). The three branches correspond to profiles identified in Fig.~\ref{fig:16}(b).The crossing point is at $\hat{T}_\mathrm{t}\simeq0.207$.}
\label{fig:17}
\end{figure}

The stability status of the three solutions is most evident in Fig.~\ref{fig:17}, which shows the free energy versus temperature.
Near and below $\hat{T}_\mathrm{H}$, the solution $\rho_\mathrm{g}$ is stable and the solution $\rho_\mathrm{d}$ is metastable. 
At $\hat{T}_\mathrm{t}$ the stability status between the solutions switches.
Here the free energies associated with the profiles $\rho_\mathrm{g}$ and $\rho_\mathrm{d}$ cross each other while the unstable profile $\rho_\mathrm{i}$ has a higher free energy.
Below $\hat{T}_\mathrm{t}$ the stable profile is $\rho_\mathrm{d}$. 
It will gradually evolve into the $T=0$ profile analyzed earlier.
The metastable solution $\rho_\mathrm{g}$ and the unstable solution $\rho_\mathrm{i}$ merge at $\hat{T}_\mathrm{L}$, where both disappear.

Notice the similarity to and difference from the MB gas.
Cooling precipitates an abrupt change in both cases. 
In the MB case that change is a gravitational collapse.
In the FD case, it is a partial collapse, arrested midway by the repulsive short-range interaction, which is rooted in the exclusion principle.

It is tempting to identify the temperature $\hat{T}_\mathrm{t}$ as the point of a first-order transition and the temperatures $\hat{T}_\mathrm{H}$, $\hat{T}_\mathrm{L}$ as spinodal points.
The significance of $\hat{T}_\mathrm{t}$ is doubtful for two resons stated below.  
The values $\hat{T}_\mathrm{H}$, $\hat{T}_\mathrm{L}$, on the other hand,
correspond to 
points of zero slope in the caloric curve.
Poincar\'e turning point criterion \cite{poincare} identifies them as points of
mechanical
instability associated with a canonical ensemble (Sec.~\ref{sec:caloric}).
Decreasing the value of $\hat{R}$ within regime (ii) toward $\hat{R}_\mathrm{c}\simeq2.655$ makes the values of $\hat{T}_\mathrm{H}$ and $\hat{T}_\mathrm{L}$ move closer together and merge at the border to regime (i).

Chavanis \cite{lifetime} pointed out that the lifetime of metastable states such as investigated here are extremely long.
For all practical purposes, they can be treated as stable macrostates.
Processes that require the transport of matter over significant energy barriers across large distances are very slow.
This is the first reason that undermines the significance of $\hat{T}_\mathrm{t}$.
The second reason is that first-order transitions at constant $\hat{T}$ are only generic in homogeneous systems.
No quasi-static processes have yet been identified between macrostates with density profiles $\rho_\mathrm{g}$ and $\rho_\mathrm{d}$.

%
\section{Phase coexistence}\label{sec:phase coex}
%
The conclusions reached in the preceding paragraph do not rule out the coexistence of a two-phase macrostate composed of segments of different solutions of the ODE (\ref{eq:23}) with boundary conditions that satisfy the applicable stability conditions.
The very long lifetimes of metastable single-phase macrostates do not speak against this possibility.
An incipient cluster can evolve from very diverse initial configurations.
A cluster may very well settle into a two-phase macrostate if such a state has a lower free energy at a given temperature than either of the single-phase macrostates.

\subsection{Conditions for single phase boundary}\label{sec:1-pha-bou}
A density profile (in $\mathcal{D}_\sigma=\mathcal{D}=3$) with confinement at radius $\hat{R}$ and one phase boundary at radius $\hat{r}_1$ results from a pair of solutions of the ODE (\ref{eq:23}) at given $\hat{T}$.
Six specifications need to be fixed. 
The two parameters,
\begin{equation}\label{eq:500} 
\hat{r}_1,\quad \varphi\doteq\frac{m_1}{m_\mathrm{tot}},
\end{equation}
locate the phase boundary and determine the mass fraction of the inner phase.
Additionally, there are the four boundary conditions,
\begin{equation}\label{eq:501}
\hat{z}(0),\quad \hat{z}'(0),\quad \hat{z}(\hat{r}_1),\quad \hat{z}'(\hat{r}_1).
\end{equation}
The two local conditions,
\begin{equation}\label{eq:502}
\hat{z}'(0)=0,\quad \hat{z}'(\hat{r}_1)=-\frac{2}{\hat{T}}\frac{\varphi}{\hat{r}_1^2}\hat{z}(\hat{r}_1),
\end{equation}
satisfy smoothness at the center and guarantee mechanical stability at the phase boundary, whereas the two integral conditions,
\begin{equation}\label{eq:503}
\int_0^{\hat{r}_1}d\hat{r}\,\hat{r}^2\rho_\mathrm{v}(\hat{r})=\varphi,\quad
\int_{\hat{r}_1}^{\hat{R}}d\hat{r}\,\hat{r}^2\rho_\mathrm{v}(\hat{r})=1-\varphi,
\end{equation}
determine the mass fractions and guarantee that the total mass is conserved.
These four conditions give us a two-parameter family of two-phase profiles separated by one phase boundary.

\subsection{Two-phase density profile}\label{sec:2-pha-den-pro}
The phase boundary is necessarily associated with a discontinuity in density -- a step down over a distance short compared to the (rather long) length scale in use.
Our task is to search for two-phase solutions within the temperature interval $\hat{T}_\mathrm{L}<\hat{T}<\hat{T}_\mathrm{H}$, where a stable and a metastable solution are known to exist.

\begin{figure}[b]
  \begin{center}
 \includegraphics[width=.48\linewidth]{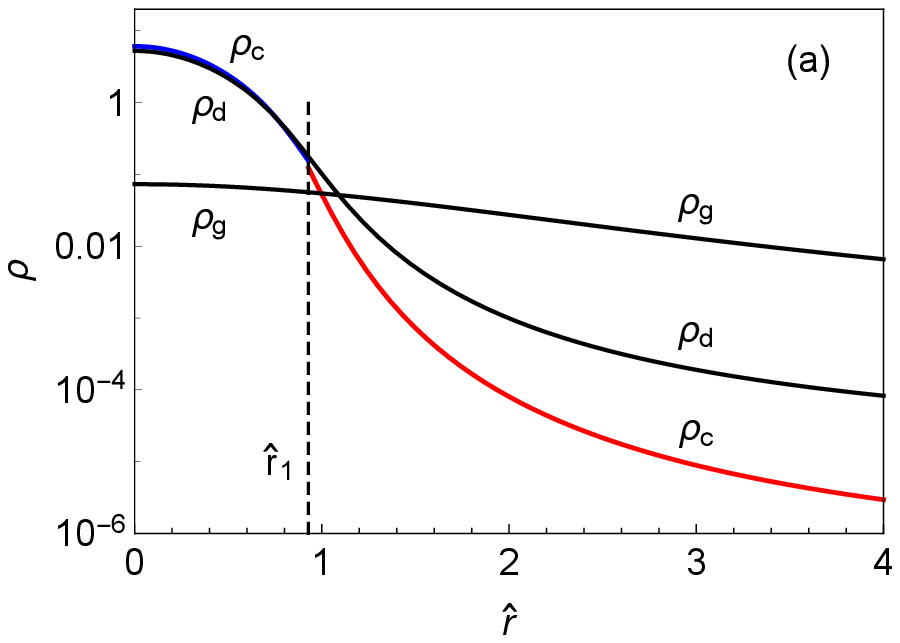}  
 \includegraphics[width=.48\linewidth]{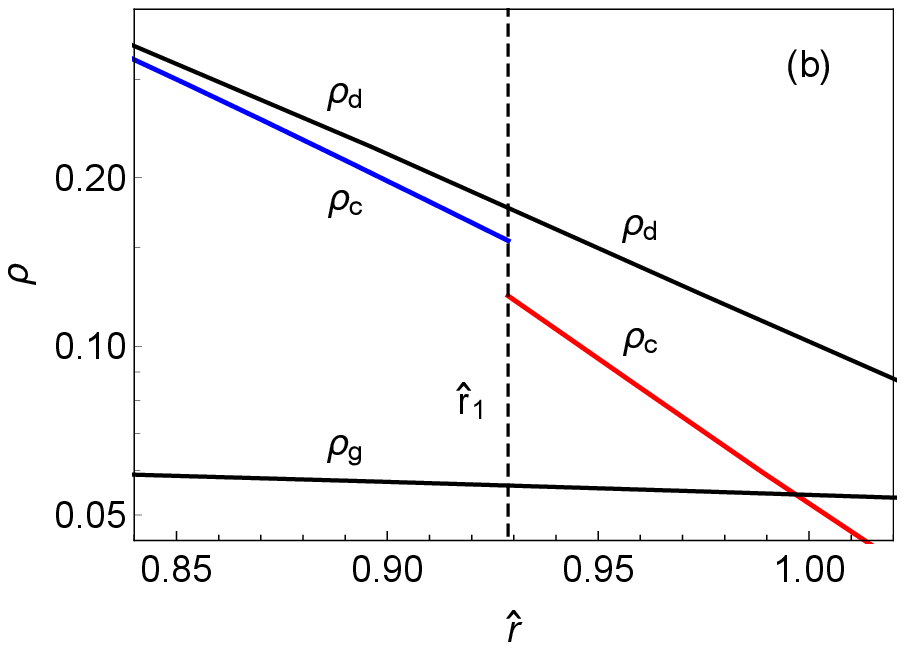}  
\end{center}
\caption{Mechanically stable density profiles at temperature $\hat{T}=0.2$ for a spherical cluster with $\hat{R}=4$. The one-phase profiles are labelled $\rho_\mathrm{g}$ and $\rho_\mathrm{d}$. The two-phase  profile $\rho_\mathrm{c}$ is discontinuous at $\hat{r}_1=0.929$. (a) Full range of scaled distance $\hat{r}$. (b) Zoomed-in view of the same data.}
  \label{fig:18}
\end{figure}

Here we merely show evidence that two-phase solutions do indeed exist and that such solutions have a lower free energy than the single-phase solutions at the same temperature.
In Fig.~\ref{fig:18} we show the data for one representative case.

The smooth one-phase profiles are again labeled $\rho_\mathrm{g}$ for the gaseous type and $\rho_\mathrm{d}$ for the core-halo type. Their free energies are,
\begin{equation}\label{eq:504}
\hat{F}_\mathrm{g}=-1.003,\quad \hat{F}_\mathrm{d}=-1.026,
\end{equation}
respectively.
The two-phase profile, labeled $\rho_\mathrm{c}$, has a step-down discontinuity at $\hat{r}_1=0.929$.
Most importantly its free energy,
\begin{equation}\label{eq:505}
\hat{F}_\mathrm{c}=-1.062,
\end{equation}
is lower than that of either single-phase profile.

Among the two-parameter family of two-phase solutions at fixed $\hat{T}$, one has the lowest-free-energy.
We expect the two-phase solution with the lowest-free-energy to merge with a type-$\rho_\mathrm{d}$ one-phase solution at $\hat{T}_\mathrm{L}$ and to merge with a type-$\rho_\mathrm{g}$ one-phase solution at $\hat{T}_\mathrm{H}$.
These expectations, if confirmed, do not yet prove that the equilibrium state in that temperature interval has a single phase boundary. 
The nontrivial mechanical stability condition may very well favor a state with more than one phase boundary.

If a first-order transition between type-$\rho_\mathrm{g}$ and
type-$\rho_\mathrm{d}$ single-phase profiles exists, then that transition takes
place over a range of temperatures, most likely the entire interval
$\hat{T}_\mathrm{L}<\hat{T}<\hat{T}_\mathrm{H}$.
Working such a scenario out in detail is computationally demanding.
It will have to include the analysis on length scale sufficiently short to resolve the structure of the phase boundary between gaseous phases.
This will be the project of a separate publication.

%
\section{Conclusion and outlook}\label{sec:concl}
%
The shape of density profiles for self-gravitating clusters of nonrelativistic FD gases depends on both the symmetry of the cluster and the dimensionality of the space. 
We have analyzed six combinations of the two attributes across the full range of temperature -- from fully degenerate clusters with compact support to the MB limit of low-density clusters.
The length scale introduced for this study turns all density profiles at given temperature into universal curves, independent of the total mass.
This length scale and the associated energy scale are equally useful for the study of BE clusters as demonstrated in a companion paper \cite{sgcbe}.

We have extended the list of analytic expressions for exact density profiles of degenerate clusters to four and identified the important specifications for all six cases, including the mass-radius relation and the cusp singularity of the density profile at the surface of the cluster.
The distinct dependence of the cluster radius on kinetic mass (smallest  fermion mass), the gravitational mass (average particle mass), and the number of fermions is potentially useful in dark matter research.

Cooling down a cluster quasistatically from high temperature changes the density profile in ways that strongly depend on the symmetry and weakly on the dimensionality.
For clusters with planar symmetry, the evolution of the density profile from the MB profile with exponential tails into a fully-degenerate compact Fermi slab is very gradual and without landmarks.
No wall confinement is necessary at any temperature.

Clusters with cylindrical symmetry, by contrast, require wall confinement against escape above a certain threshold temperature. 
Below that threshold, the density profiles have power-law tails with temperature-dependent exponents.
Cooling down a wall-confined cylindrical cluster across the threshold temperature produces an accelerated change density profile -- an incipient gravitational collapse, softly arrested midway by the implications of FD statistics.
An attribute shared by cylindrical and planar clusters is that mechanically stable macrostates are unique at any temperature.

Finite clusters with spherical symmetry, which need wall confinement at all nonzero temperatures, do not, in general, share this last attribute. 
We have identified regimes with two mechanically stable macrostates -- one thermally stable and the other metastable -- across a finite interval of temperature.
Cooling down and heating up a cluster across this interval causes mechanical instabilities at its far end, thus producing effects of hysteresis.
We have identified, inside this temperature interval, the existence of two-phase macrostate with one phase boundary, which has a lower-free-energy than either one-phase macrostate.
We have sketched a scenario of a first-order transition starting at one end of the temperature interval and reaching completion at the other end.

Investigating self-gravitating FD and BE clusters on a common length scale facilitates comparative studies such as intended here and in \cite{sgcbe}.
FD clusters and BE clusters have a common MB limit at high temperature, but evolve differently upon cooling.
BE clusters lack the robustness of FD clusters against gravitational pressure. 
Condensation is initiated at a nonzero temperature in all cases. 
The critical singularities depend on the symmetry of the cluster and dimensionality of the space \cite{sgcbe}.

A natural extension of this work will investigate a succession of relativistic effects, first the consequences of a relativistic energy-momentum relation and then the consequences of general relativity.
In the first part, currently in the works \cite{sgcrfd}, we demonstrate the
crossover of $T=0$ density profiles between universal nonrelativistic shapes and
universal ultrarelativistic shapes.
We also investigate how the mass-radius relation varies with the symmetry of the cluster and the dimensionality of the space and how the stability of spherical FD clusters depend on mass and temperature.

\appendix

%
\section{Alternative scaling convention}\label{sec:appa}
%
Here we establish, for the sake of transparency, the relations between the length scale and energy scale used in Ref.~\cite{ptdimd} and the scaling conventions introduced in Sec.~\ref{sec:scales} for this work.
We begin by stating the explicit dependence of the length scale $r_\mathrm{s}$ and the energy scale $k_\mathrm{B}T_\mathrm{s}$ used in this work as inferred from Eqs.~(\ref{eq:17}):
\begin{align}\label{eq:25}
&(k_\mathrm{B}T_\mathrm{s})^{1+\mathcal{D}/\mathcal{D}_\sigma-\mathcal{D}/2} 
=\frac{1}{2}G_\mathcal{D}\frac{\mathcal{A_D}}{\mathcal{D}_\sigma}
\left(\frac{\mathcal{A}_{\mathcal{D}_\sigma}}{\mathcal{D}_\sigma}\right)^{-2/\mathcal{D}_\sigma}
 \nonumber \\
 & \hspace{5mm}\times(2\pi\hbar^2)^{\mathcal{D}/\mathcal{D}_\sigma-\mathcal{D}/2} 
M^2m^{-\mathcal{D}/\mathcal{D}_\sigma+\mathcal{D}/2}
\tilde{N}^{2/\mathcal{D}_\sigma},
\end{align}
\begin{equation}\label{eq:26}
r_\mathrm{s}^{\mathcal{D}_\sigma}=
\frac{\mathcal{D}_\sigma}{\mathcal{A}_{\mathcal{D}_\sigma}}
(2\pi\hbar^2)^{\mathcal{D}/2}\tilde{N}\,m^{-\mathcal{D}/2}
(k_\mathrm{B}T_\mathrm{s})^{-\mathcal{D}/2}.
\end{equation}
The alternative scaled length (for $\mathcal{D}_\sigma=\mathcal{D}$) is defined as follows (in our units):
\begin{equation}\label{eq:66}
\xi\doteq\frac{r}{r_\mathrm{P}},\quad r_\mathrm{P}^2=\frac{k_\mathrm{B}T\lambda_T^\mathcal{D}}
{g_s\mathcal{A_D}G_\mathcal{D}m^2}\,\Gamma(\mathcal{D}/2).
\end{equation}
In the limit $T\to0$, the length scale $r_\mathrm{P}$ shrinks to zero in $\mathcal{D}=1$, stretches to infinity in $\mathcal{D}=3$, and is $T$-independent in $\mathcal{D}=2$. 
Its relation to $r_\mathrm{s}$ from (\ref{eq:17}) is
\begin{equation}\label{eq:68}
\left(\frac{r_\mathrm{P}}{r_\mathrm{s}}\right)^2=
\frac{\Gamma(\mathcal{D}/2)}{2\mathcal{D}g_s}\,\hat{T}^{1-\mathcal{D}/2}.
\end{equation}

The bridge between our ODE (\ref{eq:11}) for the fugacity $z(r)$ and the corresponding ODE arrived at in Ref.~\cite{ptdimd} is spanned as follows.
We write,
\begin{equation}\label{eq:69}
z_\xi(\xi)\doteq z(r),\quad z'_\xi=z'r_\mathrm{P},\quad z''_\xi=z''r_\mathrm{P}^2,
\end{equation}
which transcribes (\ref{eq:11}) into
\begin{equation}\label{eq:70}
\frac{z_\xi''}{z_\xi}+\frac{\mathcal{D}-1}{\xi}\frac{z_\xi'}{z_\xi}
-\left(\frac{z_\xi'}{z_\xi}\right)^2+\Gamma(\mathcal{D}/2)f_{\mathcal{D}/2}(z_\xi)=0.
\end{equation}
With the relation,
\begin{equation}\label{eq:71}
z_\xi^{-1}(\xi)=k\,e^{\psi(\xi)},
\end{equation}
between the fugacity and the potential $\psi(\xi)$ the ODE (\ref{eq:70}) becomes
\begin{equation}\label{eq:72}
\frac{1}{\xi^{\mathcal{D}-1}}\frac{d}{d\xi}
\left(\xi^{\mathcal{D}-1}\frac{d\psi}{d\xi}\right)
=I_{\mathcal{D}/2-1}\big(ke^{\psi(\xi)}\big), 
\end{equation}
where $I_n(t)=\Gamma(n-1)f_{n+1}(t^{-1})$.
Both scaling conventions produce one-parameter families of solutions.
The parameter only enters one of the boundary conditions.
For (\ref{eq:72}) that parameter is $k$ and for (\ref{eq:70}) it is $z_\xi(0)$.

Either parameter, $k$ or $z_\xi(0)$, contains several physical quantities that we might wish to vary separately: the confining radius $R$, the temperature $T$, and the number $N$ of particles (or the total mass $m_\mathrm{tot}=NM$).
In \cite{ptdimd} the parameter $k$ is split into two (dimensionless) parts.
One is the scaled radius of confinement,
\begin{equation}\label{eq:73}
\alpha\doteq\frac{R}{r_\mathrm{P}},
\end{equation}
and the other the degeneracy parameter,
\begin{equation}\label{eq:75}
\mu\doteq\eta_0\mathcal{A}^2_\mathcal{D}2^{\mathcal{D}/2-1}G_\mathcal{D}^{\mathcal{D}/2}
m_\mathrm{tot}^{\mathcal{D}/2-1}R^{\mathcal{D}(4-\mathcal{D})/2}.
\end{equation}
The energy scale used in \cite{ptdimd} can be stated as follows:
\begin{equation}\label{eq:76}
\eta^{-1}\doteq\frac{k_\mathrm{B}T}{k_\mathrm{B}T_\mathrm{P}},\quad
k_\mathrm{B}T_\mathrm{P}=\frac{G_\mathcal{D}mm_\mathrm{tot}}{R^{\mathcal{D}-2}},
\end{equation}
Its relation to $k_\mathrm{B}T_\mathrm{s}$ from (\ref{eq:17}) and (\ref{eq:76}) is
\begin{equation}\label{eq:77}
\frac{T_\mathrm{P}}{T_\mathrm{s}}=2\hat{R}^{2-\mathcal{D}}.
\end{equation}
The scaled temperatures are related via
\begin{equation}\label{eq:78}
\eta^{-1}=\frac{1}{2}\hat{R}^{\mathcal{D}-2}\hat{T}.
\end{equation}
The scaled radius of confinement $\alpha$ is $T$-dependent whereas $\hat{R}$ is not.
The two are related as follows:
\begin{equation}\label{eq:79}
\alpha=\sqrt{\frac{2\mathcal{D}g_s}{\Gamma(\mathcal{D}/2)}}
\hat{T}^{(\mathcal{D}/2-1)/2}\hat{R}.
\end{equation}
It is useful to express the degeneracy parameter $\mu$ in terms of the scaled radius of confinement $\alpha$ and the scaled inverse temperature $\eta$:
\begin{equation}\label{eq:80}
\mu=\alpha^2\eta^{\mathcal{D}/2-1}.
\end{equation}
The relation between $\mu$ and $\hat{R}$ then follows directly:
\begin{equation}\label{eq:81}
\mu=2^{\mathcal{D}/2}\frac{\mathcal{D}g_s}{\Gamma(\mathcal{D}/2)}\,
\hat{R}^{2\mathcal{D}-\mathcal{D}^2/2}.
\end{equation}

Caloric curves can be produced alternatively by keeping $\mu$ fixed and varying $\alpha$ (as done in Ref.~\cite{ptdimd}) or by keeping $\hat{R}$ fixed and varying $\hat{T}$ (as done in Sec.~\ref{sec:caloric}). 
The two sets are not identical, but there is a one-on-one correspondence between maxima, minima, and locations of infinite slope. 
Note that the scale $T_s$ depends on $\hbar$ but not on $R$; it is adapted to the large domain limit $R\rightarrow \infty$. 
By contrast, the scale $T_P$ depend on $R$ but not on $\hbar$; it is adapted to the classical limit $\hbar\rightarrow 0$.

%
\section{Virial theorem}\label{sec:appb}
%
Here we develop an expression for the virial theorem pertaining to clusters with $\mathcal{D}_\sigma<\mathcal{D}$ in generalization to the results for $\mathcal{D}_\sigma=\mathcal{D}$ previously established by Chavanis and Sire \cite{CS04}. 
We begin with the definition of the virial for the gravitational force,
\begin{equation}
\label{eq:C1}
\mathcal{V}_{\mathcal{D}_\sigma,\mathcal{D}}\doteq m\int  d^\mathcal{D}r\,\rho_\mathrm{v}
\,\mathbf{r}\cdot \nabla\Phi.
\end{equation}
The gravitational potential $\Phi$ is inferred from (\ref{eq:8}) via 
\begin{equation}
\label{eq:C2}
\frac{d}{dr}\Phi(r) =-g(r)=\frac{\mathcal{A}_{\mathcal{D}}}{\mathcal{A}_{\mathcal{D}_\sigma}}\frac{G_\mathcal{D}\tilde{M}(r)}{r^{\mathcal{D}_\sigma-1}},
\end{equation}
where 
\begin{equation}
\label{eq:C3}
\tilde{M}(r)\doteq \frac{M(r)}{L^{\mathcal{D}-\mathcal{D}_\sigma}}=m\mathcal{A}_{\mathcal{D}_\sigma}\int_0^r dr' ~r'^{\mathcal{D}_\sigma-1}\rho_\mathrm{v}(r')~~.
\end{equation} 
We can thus simplify the integral in (\ref{eq:C1}):
\begin{equation}
\label{C4}
\mathcal{V}_{\mathcal{D}_\sigma,\mathcal{D}}=\frac{\mathcal{A}_{\mathcal{D}}}{\mathcal{A}_{\mathcal{D}_\sigma}}\frac{G_\mathcal{D}L^{\mathcal{D}-\mathcal{D}_\sigma}}{2}\int_0^R 
\frac{dr}{r^{\mathcal{D}_\sigma-2}}\frac{d}{dr}\big[\tilde{M}(r)\big]^2.
\end{equation}
For cases with $\mathcal{D}_\sigma=2$ the integral is simple and, alas, useless for our purpose:
\begin{equation}
\label{eq:C5}
\mathcal{V}_{2,\mathcal{D}}=\frac{\mathcal{A}_{\mathcal{D}}}{4\pi}G_\mathcal{D}\tilde{m}_\mathrm{tot}m_\mathrm{tot},
\end{equation}
where $\tilde{m}_\mathrm{tot}\doteq m_\mathrm{tot}/L^{\mathcal{D}-\mathcal{D}_\sigma}$.
For cases with $\mathcal{D}_\sigma\neq2$, an integration by parts brings the  virial into the form,
\begin{align}\label{eq:C6}
\mathcal{V}_{\mathcal{D}_\sigma,\mathcal{D}}
= &\frac{\mathcal{A}_{\mathcal{D}}}{\mathcal{A}_{\mathcal{D}_\sigma}}
\frac{G_\mathcal{D}}{2}L^{\mathcal{D}-\mathcal{D}_\sigma} \nonumber \\
&\hspace{-5mm}\times \left[ \frac{\tilde{m}_\mathrm{tot}^2}{R^{\mathcal{D}_\sigma-2}} 
+(\mathcal{D}_\sigma-2)\int_0^Rdr ~\frac{[\tilde{M}(r)]^2}{r^{\mathcal{D}_\sigma-1}}\right].
\end{align}
With \eqref{eq:C2} we can relate the virial more directly to the gravitational
potential:
\begin{align}\label{eq:C7}
\mathcal{V}_{\mathcal{D}_\sigma,\mathcal{D}}
&=\frac{\mathcal{A}_\mathcal{D}G_\mathcal{D}L^{\mathcal{D}-\mathcal{D}_\sigma}\tilde{m}_\mathrm{tot}^2}{2\mathcal{A}_{\mathcal{D}_\sigma}R^{\mathcal{D}_\sigma-2}} \\ \nonumber 
&+(\mathcal{D}_\sigma-2)\frac{\mathcal{A}_{\mathcal{D}_\sigma}L^{\mathcal{D}-\mathcal{D}_\sigma}}{2\mathcal{A}_\mathcal{D}G_\mathcal{D}}\int_0^R dr\,r^{\mathcal{D}_\sigma-1}\left[\frac{d\Phi}{dr}\right]^2.
\end{align}

The next thread relates the gravitational potential energy $W$ to the same $\Phi(r)$:
\begin{align}\label{eq:C8}
W&=\frac{m}{2}\int d^\mathcal{D}r\, \rho_\mathrm{v}\Phi
=\frac{1}{2~\mathcal{A}_\mathcal{D}G_\mathcal{D}}\int d^\mathcal{D}r\,\Phi \nabla^2 \Phi
 \nonumber \\
&=\frac{\mathcal{A}_{\mathcal{D}_\sigma}L^{\mathcal{D}-\mathcal{D}_\sigma}}{2~\mathcal{A}_\mathcal{D}G_\mathcal{D}}\left[ R^{\mathcal{D}_\sigma-1}\Phi(R)\frac{d}{dr}\Phi(R)
 \right. \nonumber \\
& \hspace{25mm}\left.-\int_0^R dr~r^{\mathcal{D}_\sigma-1}\left[\frac{d \Phi}{dr} \right]^2  \right],
\end{align}
where we used the Newton-Poisson equation, $\nabla^2
\Phi=\mathcal{A}_\mathcal{D} G_\mathcal{D} m\rho_\mathrm{v}$, in the first step,
and integrated by parts in the second. 
We can simplify the first term in the last expression by using \eqref{eq:C2}, evaluated at $R$ with $\tilde{M}(R)=\tilde{m}_\mathrm{tot}$.

Convenient choices for the integration constants set ${\Phi(0)=0}$ in $\mathcal{D}_\sigma=1$, $\Phi(R)=0$
in $\mathcal{D}_\sigma=2$, and $\Phi(\infty)=0$ in $\mathcal{D}_\sigma>2$.
The expressions, valid for $r\leq R$, read
\begin{subequations}\label{eq:C10}
\begin{align}
\Phi(r)&=-\frac{1}{(\mathcal{D}_\sigma-2)}\frac{\mathcal{A}_\mathcal{D}G_\mathcal{D}\tilde{m}_\mathrm{tot}}{\mathcal{A}_{\mathcal{D}_\sigma}r^{\mathcal{D}_\sigma-2}},
\quad :~ \mathcal{D}_\sigma\neq2, \\
\Phi(r)&=\frac{\mathcal{A}_\mathcal{D}G_\mathcal{D}\tilde{M}}{2\pi}\ln\left(\frac{r}{R}\right)
\quad :~ \mathcal{D}_\sigma=2.
\end{align}
\end{subequations}
With these substitutions \eqref{eq:C8} becomes,
\begin{subequations}\label{eq:C11}
\begin{align}
W=&-\frac{\mathcal{A}_{\mathcal{D}}L^{\mathcal{D}-\mathcal{D}_\sigma}}
{2\,\mathcal{A}_{\mathcal{D}_\sigma}(\mathcal{D}_\sigma-2)}\frac{G_\mathcal{D}\tilde{m}_\mathrm{tot}^2}{ R^{\mathcal{D}_\sigma-2}}  \\ \nonumber
&-\frac{\mathcal{A}_{\mathcal{D}_\sigma}L^{\mathcal{D}-\mathcal{D}_\sigma}}{2\,\mathcal{A}_\mathcal{D} G_\mathcal{D}}\int_0^R dr\,r^{\mathcal{D}_\sigma-1}\left[\frac{d \Phi}{dr} \right]^2
\quad :~ \mathcal{D}_\sigma\neq2, \\
W&=-\frac{\pi L^{\mathcal{D}-2}}{\mathcal{A}_\mathcal{D} G_\mathcal{D}}\int_0^R dr~r\left[\frac{d \Phi}{dr} \right]^2 \quad :~ \mathcal{D}_\sigma=2.
\end{align}
\end{subequations}
Comparison of \eqref{eq:C11} with \eqref{eq:C7} yields the important intermediate result,
\begin{equation}\label{eq:C12}
\mathcal{V}_{\mathcal{D}_\sigma, \mathcal{D}}=-(\mathcal{D}_\sigma-2)W.
\end{equation}
Next we establish the relationship between the virial $\mathcal{V}_{\mathcal{D}_\sigma,\mathcal{D}}$ and the kinetic energy $U$.
We start from the condition (\ref{eq:6}) of mechanical equilibrium,
\begin{equation}\label{eq:C13}
\nabla p =-m \rho_\mathrm{v}\nabla \Phi,
\end{equation}
which permits the virial \eqref{eq:C1} to be rendered in the form,
\begin{equation}\label{eq:C14}
\mathcal{V}_{\mathcal{D}_\sigma,\mathcal{D}}=\int d^\mathcal{D}r\,\mathbf{r}\cdot\nabla p,
\end{equation}
and, after an integration by parts, in the form,
\begin{align}\label{eq:C15}
\mathcal{V}_{\mathcal{D}_\sigma,\mathcal{D}}
&=-\mathcal{A}_{\mathcal{D}_\sigma}L^{\mathcal{D}-\mathcal{D}_\sigma}R^{\mathcal{D}_\sigma}p(R) \nonumber \\
&+\mathcal{D}_\sigma \mathcal{A}_{\mathcal{D}_\sigma}L^{\mathcal{D}-\mathcal{D}_\sigma} \int_0^Rdr\,r^{\mathcal{D}_\sigma-1}p(r). 
\end{align}
The second term in this expression is related to the kinetic energy via the EOS
(\ref{eq:1}), which thus produces the second intermediate result (still for
$\mathcal{D}_\sigma \neq 2$),
\begin{equation}
\label{eq:C16}
2\frac{\mathcal{D}_\sigma}{\mathcal{D}}U-\mathcal{V}_{\mathcal{D}_\sigma,\mathcal{D}}=\mathcal{A}_{\mathcal{D}_\sigma}L^{\mathcal{D}-\mathcal{D}_\sigma}R^{\mathcal{D}_\sigma}p(R).
\end{equation}
The combination of \eqref{eq:C12} and \eqref{eq:C16}, which relates the
potential energy and the kinetic energy for self-gravitating FD clusters with
planar symmetry $(\mathcal{D}=1)$ or spherical symmetry $(\mathcal{D}=3)$, thus
constitutes the virial theorem in the general form required for this study:
\begin{equation}
\label{eq:C18}
2\frac{\mathcal{D}_\sigma}{\mathcal{D}}U+(\mathcal{D}_\sigma-2)W
=\mathcal{A}_{\mathcal{D}_\sigma}L^{\mathcal{D}-\mathcal{D}_\sigma}R^{\mathcal{D}_\sigma}p(R).
\end{equation}
For clusters with cylindrical symmetry $(\mathcal{D}_\sigma=2)$, the combination of \eqref{eq:C5} and \eqref{eq:C16} produce the identity,
\begin{equation}\label{eq:C17}
\frac{4}{\mathcal{D}}U-\frac{\mathcal{A}_{\mathcal{D}}}{4\pi}G_\mathcal{D}\tilde{m}_\mathrm{tot}m_\mathrm{tot}=2\pi L^{\mathcal{D}-2}R^2p(R).
\end{equation}

From the virial theorem, we can obtain explicit exact results in special cases: 

(i) If we consider $\mathcal{D}_\sigma=1$, $p(R)=0$ (valid in an infinite
domain or when the density profile is a Dirac distribution),
and the MB statistics, from the relations $E=U+W$,
$2\frac{1}{\mathcal{D}}U-W =0$ and $U=\frac{\mathcal{D}}{2}Nk_B T$, we find
that the caloric curve is given by
\begin{equation}
E=\left (\frac{\mathcal{D}}{2}+1\right )Nk_B T.
\end{equation}

(ii) If we consider $\mathcal{D}_\sigma=2$, $p(R)=0$ (valid in an infinite
domain or when the density profile is a Dirac distribution), and the MB
statistics, from the relations 
$\frac{4}{\mathcal{D}}U-\frac{\mathcal{A}_{\mathcal{D}}}{4\pi}G_\mathcal{
D}\tilde {m}_\mathrm{tot}m_\mathrm{tot}=0$ and
$U=\frac{\mathcal{D}}{2}Nk_B T$, we find that the temperature is given by
\begin{equation}
k_B T=\frac{\mathcal{A}_{\mathcal{D}}}{8\pi N}G_\mathcal{
D}\tilde {m}_\mathrm{tot}m_\mathrm{tot}.
\end{equation}

%
\section{Polytropes}\label{sec:appd}
%
Gaseous polytropes with index $n$ are characterized by a pure power-law dependence of pressure on density:
\begin{equation}
\label{eq:D1}
p\propto \rho^{\gamma},\quad \gamma= 1+\frac{1}{n}.
\end{equation}

Internal energy expressions for polytropes with index $n=\mathcal{D}/2$ are a
useful way to characterize FD clusters in $\mathcal{D}$ dimensions at $T=0$.
Following the strategy of Ref.~\cite{CS04}, but including clusters with 
$\mathcal{D}_\sigma<\mathcal{D}$, we begin by extracting from the asymptotics
carried out in Sec.~\ref{sec:degen} the relation between pressure and density,
\begin{equation}
\label{eq:D2}
p(r)=\left(\frac{\Lambda^{\mathcal{D}/2} \Gamma(\mathcal{D}/2+1)}{g_s}\right)^{2/\mathcal{D}}\frac{\rho_\mathrm{v}(r)^{1+2/\mathcal{D}}}{(\mathcal{D}/2+1)},
\end{equation}
which has the form (\ref{eq:D1}) 
and can be rendered as follows:
\begin{equation}
\label{eq:D4}
\frac{1}{\rho_\mathrm{v}(r)}\frac{d}{dr}p(r)=(\mathcal{D}/2+1)\frac{d}{dr}\left( \frac{p(r)}{\rho_\mathrm{v}(r)}\right).
\end{equation}
Combining it with the condition (\ref{eq:C13}) of mechanical equilibrium, we 
obtain
\begin{equation}
\label{eq:D5}
-\frac{(\mathcal{D}/2+1)}{m}\frac{d}{dr}\left( \frac{p(r)}{\rho_\mathrm{v}(r)}\right)=\frac{d}{dr}\Phi(r).
\end{equation}
The profile of the gravitational potential thus becomes,
\begin{equation}
\label{eq:D6}
\Phi(r)=\Phi(R)-\frac{(\mathcal{D}/2+1)}{m}\left( \frac{p(r)}{\rho_\mathrm{v}(r)}-\frac{p(R)}{\rho_\mathrm{v}(R)}\right)~~.
\end{equation}
Expressing the internal energy $E=U+W$ and the number of particles $N$ via the integrals,
\begin{equation}
\label{eq:D9}
E=\frac{\mathcal{D}}{2}\int d^\mathcal{D}r p+\frac{m}{2}\int d^\mathcal{D}r \rho_\mathrm{v}\Phi, \quad N=\int d^\mathcal{D}r \rho_\mathrm{v},
\end{equation}
we can cast the potential energy into the form,
\begin{equation}
\label{eq:D10}
W=-\frac{(\mathcal{D}/2+1)}{\mathcal{D}}U+\frac{(\mathcal{D}/2+1)N}{2}\frac{p(R)
}{\rho_\mathrm{v}(R)}+\frac{Nm}{2}\Phi(R).
\end{equation}
Next we eliminate the kinetic energy $U$ from \eqref{eq:D10} by invoking the
virial theorem, i.e. 
relations (\ref{eq:C18}) for ${\mathcal{D}_\sigma\neq2}$ and (\ref{eq:C17}) for
${\mathcal{D}_\sigma=2}$.
We thus obtain,
\begin{subequations}\label{eq:D13}
\begin{align}\label{eq:D13a}
W&=\frac{2\mathcal{D}_\sigma }{2\mathcal{D}_\sigma -(\mathcal{D}/2+1)(\mathcal{D}_\sigma-2)} \nonumber \\
&\times\bigg[-\frac{\mathcal{A}_\mathcal{D}G_\mathcal{D}\tilde{m}_\mathrm{tot}m_\mathrm{tot}}{2\mathcal{A}_{\mathcal{D}_\sigma}(\mathcal{D}_\sigma-2)} R^{2-\mathcal{D}_\sigma}
+\frac{N(\mathcal{D}/2+1)}{2}\nonumber \frac{p(R)}{\rho_\mathrm{v}(R)}\\
&~~~~~~ -\frac{(\mathcal{D}/2+1)}{2\mathcal{D}_\sigma}\mathcal{A}_{\mathcal{D}_\sigma}L^{\mathcal{D}-\mathcal{D}_\sigma}R^{\mathcal{D}_\sigma}p(R)\bigg], \\
\label{eq:D13b}
W&=\frac{N(\mathcal{D}/2+1)}{2}\frac{p(R)}{\rho_\mathrm{v}(R)}
\\
&-\frac{(\mathcal{D}/2+1)}{4}\bigg[\frac{\mathcal{A}_\mathcal{D}}{4\pi}G_\mathcal{D}\tilde{m}_\mathrm{tot}m_\mathrm{tot} +2\pi L^{\mathcal{D}-2}R^2p(R)\bigg], \nonumber 
\end{align}
\end{subequations}
for $\mathcal{D}_\sigma\neq2$ and $\mathcal{D}_\sigma=2$, respectively.
We have used
$\Phi(R)=-\mathcal{A}_\mathcal{D}G_\mathcal{D}\tilde{M}/[\mathcal{A}_{\mathcal{D
}_\sigma}(\mathcal{D}_\sigma-2)R^{\mathcal{D}_\sigma-2}]$ for
$\mathcal{D}_\sigma\neq 2$ and $\Phi(R)=0$ when $\mathcal{D}_\sigma= 2$. 
Using (\ref{eq:C18}), (\ref{eq:C17}), and \eqref{eq:D13} we thus simplify the
internal energy expressions for $\mathcal{D}_\sigma\neq2$ and
$\mathcal{D}_\sigma=2$ into
\begin{align}\label{eq:D14}
E&=\frac{\mathcal{D}}{2\mathcal{D}_\sigma} \mathcal{A}_{\mathcal{D}_\sigma}L^{\mathcal{D}-\mathcal{D}_\sigma}R^{\mathcal{D}_\sigma}p(R)
+\left(1-\frac{\mathcal{D}(\mathcal{D}_\sigma-2)}{2\mathcal{D}_\sigma}
\right)W,
 \nonumber \\
E&=\frac{(\mathcal{D}/2-1)}{4}\left[\frac{\mathcal{A}_\mathcal{D}}{4\pi}G_\mathcal{D}
\tilde{m}_\mathrm{tot}m_\mathrm{tot}
+2\pi L^{\mathcal{D}-2}R^2p(R) \right]\nonumber\\
&~~~~~~~~~~~~~~~~~~~ + \frac{N(\mathcal{D}/2+1)}{2}\frac{p(R)}{\rho_\mathrm{v}(R)}, 
\end{align}
respectively.
The above expression for $\mathcal{D}_\sigma=2$ is only useful for incomplete polytropes, where $p(R)/\rho_\mathrm{v}(R)\neq 0$, or complete polytropes with $\mathcal{D}\neq 2$. 
For complete polytropes, including those where
$\mathcal{D}_\sigma=\mathcal{D}=2$, we derive an 
alternate expression for the potential energy. Replacing $R$
by $r_0$, the radius marking the surface of the
cluster, and using $p(r_0)/\rho_\mathrm{v}(r_0)=0$, eq. \eqref{eq:D13} reduces
to the
Betti-Ritter formula \cite{chandrabook}:
\begin{align}\label{eq:D15a}
W&=\frac{-\mathcal{D}_\sigma }{2\mathcal{D}_\sigma -(\mathcal{D}/2+1)(\mathcal{D}_\sigma-2)}\frac{\mathcal{A}_\mathcal{D}G_\mathcal{D}\tilde{M}M}{\mathcal{A}_{\mathcal{D}_\sigma}(\mathcal{D}_\sigma-2)}r_0^{2-\mathcal{D}_\sigma}, \nonumber\\
W&=\frac{\mathcal{A}_\mathcal{D}G_\mathcal{D}\tilde{m}_\mathrm{tot}m_\mathrm{tot}}{4\pi}\ln\left(\frac{r_0}{R}\right)-(\mathcal{D}/2+1)\frac{\mathcal{A}_\mathcal{D}G_\mathcal{D}\tilde{M}M}{16\pi},
\end{align}
for $\mathcal{D}_\sigma\neq2$ and $\mathcal{D}_\sigma=2$, respectively.
The energy for the complete polytrope can now be written in the form,
\begin{align}\label{eq:D16}
E&=-\frac{1}{2}~\frac{2\mathcal{D}_\sigma -\mathcal{D}(\mathcal{D}_\sigma-2)}{2\mathcal{D}_\sigma -(\mathcal{D}/2+1)(\mathcal{D}_\sigma-2)}~\frac{\mathcal{A}_\mathcal{D}G_\mathcal{D}\tilde{m}_\mathrm{tot}m_\mathrm{tot}}{\mathcal{A}_{\mathcal{D}_\sigma}(\mathcal{D}_\sigma-2)r_0^{\mathcal{D}_\sigma-2}},
\nonumber\\
E&=\frac{\mathcal{A}_\mathcal{D}G_\mathcal{D}\tilde{m}_\mathrm{tot}m_\mathrm{tot}}{4\pi}\ln\left(\frac{r_0}{R}\right) \nonumber \\
&\hspace{20mm}+(\mathcal{D}/2-1)\frac{\mathcal{A}_\mathcal{D}G_\mathcal{D}\tilde{m}_\mathrm{tot}m_\mathrm{tot}}{16\pi},
\end{align}
where we have used $E=U+W$ and eliminated the kinetic energy by means of the virial theorem.




\begin{thebibliography}{100}


\bibitem{fowler}
{\small R.H. Fowler, {\it On dense matter},
MNRAS, {\bf 87}, 114 (1926)}

\bibitem{chandrabook}  {
\small  S. Chandrasekhar,
{\it An Introduction to the Theory of Stellar Structure} (University of Chicago
Press, 1939)}

\bibitem{emden}  
{\small R. Emden, {\it Gaskugeln} (Teubner Verlag, Leipzig, 
1907)}

\bibitem{stoner29}
{\small E.C. Stoner, {\it The limiting density in white dwarf stars} Phil. Mag.
{\bf 7}, 63
(1929)}

\bibitem{milne}
{\small E.A. Milne, {\it The analysis of stellar structure},
MNRAS {\bf 91}, 4
(1930)}

\bibitem{chandra31nr}
{\small S. Chandrasekhar, {\it The density of white dwarf stars} Phil.
Mag. {\bf 11}, 592
(1931)}

\bibitem{htf}
{\small P. Hertel and W. Thirring, {\it Free energy of gravitating fermions},
Commun. Math. Phys. {\bf 
24}, 22 (1971)}


\bibitem{ht}
{\small P. Hertel and W. Thirring, Thermodynamic Instability
of a System of Gravitating Fermions. In: H.P. D\"urr (Ed.): {\it Quanten und
Felder} (Brauschweig: Vieweg 1971)}


\bibitem{antonov} {\small V.A. Antonov, Vest. Leningr. Gos. Univ. {\bf 7}, 135
(1962); Translation in  IAU Symposium {\bf 113}, 525 (1985)}


\bibitem{lbw}
{\small  D. Lynden-Bell, R. Wood, {\it The gravo-thermal catastrophe in
isothermal spheres and the onset of red-giant structure for stellar
systems}, Mon. Not. R. Astron. Soc.
{\bf 138}, 495 (1968)}


\bibitem{bvn}
{\small N. Bilic, R.D. Viollier, {\it Gravitational phase transition of
fermionic matter}, Phys. Lett. B {\bf  408}, 75
(1997)}

\bibitem{pt}  
{\small P.H. Chavanis, {\it Phase transitions in self-gravitating systems.
Self-gravitating fermions and hard spheres models}, Phys. Rev. E {\bf 65},
056123 (2002)}


\bibitem{ijmpb}  {\small P.H. Chavanis, {\it Phase transitions in
self-gravitating systems}, Int. J. Mod. Phys. B {\bf 20}, 3113
(2006) }



\bibitem{lifetime}  
{\small P.H. Chavanis, {\it On the lifetime of metastable states in
self-gravitating systems}, Astron. Astrophys. {\bf 432}, 117
(2005)}


\bibitem{clm2}
{\small P.H. Chavanis, M. Lemou, F. M\'ehats, {\it Models of dark matter halos
based on statistical mechanics : The fermionic King model}, Phys. Rev. D {\bf
92}, 123527 (2015)}


\bibitem{Chav07}
P.-H. Chavanis,
\emph{White dwarf stars in $\mathcal{D}$ dimensions},
Phys. Rev. D \textbf{76}, 023004 (2007).


\bibitem{ptdimd}
{\small P.H. Chavanis, {\it Statistical mechanics and thermodynamic limit of
self-gravitating fermions in $D$ dimensions}, Phys. Rev. E  {\bf 69},
066126 (2004)}


\bibitem{ehrenfest}
{\small P. Ehrenfest, {\it In what way does it become manifest in the
fundamental laws of physics that space has three dimensions?} Proc. Amst. Acad.
{\bf 20}, 200
(1917)} 



\bibitem{shapiroteukolsky} 
{\small  S.L. Shapiro, S.A.
Teukolsky
{\it Black Holes, White Dwarfs, and Neutron Stars} (Wiley Interscience, 1983)}





\bibitem{lb}  
{\small D. Lynden-Bell, {\it Statistical mechanics of violent relaxation in
stellar systems}, Mon. Not. Roy. Astr. Soc.  {\bf 136}, 101
(1967)}



\bibitem{chavmnras}
{\small P.H. Chavanis, {\it On the coarse-grained evolution of collisionless
stellar systems}, Mon. Not. R. Astron. Soc.
{\bf 300}, 981 (1998)}




\bibitem{csmnras}
{\small P.H. Chavanis, J. Sommeria, {\it Degenerate equilibrium states of
collisionless stellar systems}, Mon. Not. R. Astron. Soc.
{\bf 296}, 569 (1998)}


\bibitem{bmtv}  
{\small N. Bilic, F. Munyaneza, G.B. Tupper, R.D. Viollier, {\it The dynamics of
stars near Sgr A* and dark matter at the center and in the halo of the Galaxy},
Prog. Part. Nucl. Phys.  {\bf 48}, 291 (2002)}

\bibitem{btv}
{\small N. Bilic, G.B. Tupper, R.D. Viollier, {\it Dark Matter in the Galaxy},
Lect. Notes Phys.
{\bf  616}, 24 (2003)}


\bibitem{rar}
{\small R. Ruffini, C.R. Arg\"uelles, J.A. Rueda, 
{\it On the core-halo distribution of dark matter in galaxies},
Mon. Not. R.
Astron. Soc. {\bf 451}, 622 (2015)}

\bibitem{rarnew}
{\small C.R. Arg\"uelles, M.I. D\'iaz, A. Krut, R. Yunis, {\it On the formation
and stability of fermionic dark matter haloes in a cosmological framework}, Mon.
Not. R. astr. Soc. {\bf 502}, 4227 (2021)}

\bibitem{mcmh}
{\small P.H. Chavanis, {\it Derivation of the core mass-halo mass relation of
fermionic and bosonic dark matter halos from an effective thermodynamical
model}, Phys. Rev. D {\bf 100}, 123506 (2019)}


\bibitem{modeldm}
{\small P.H. Chavanis, {\it Predictive model of fermionic dark matter halos with
a quantum core 
and an isothermal atmosphere }
(unpublished)}




\bibitem{sgcbe}
M. Kirejczyk, G. M\"uller, and P.-H. Chavanis,
{\it Self-gravitating Bose-Einstein gas with planar, cylindrical, or spherical
symmetry: density profiles and onset of condensation}
(unpublished).

\bibitem{sgcrfd}
M. Kirejczyk, G. M\"uller, and P.-H. Chavanis,
{\it Relativistic effects in self-gravitating clusters of Fermi-Dirac gas \\
with planar, cylindrical, or spherical symmetry}
(unpublished).


\bibitem{CS04}
P.-H. Chavanis and C. Sire,
\emph{Anomalous diffusion and collapse of self-gravitating Langevin particles in
$D$ dimensions},
Phys. Rev. E \textbf{69}, 016116 (2004).


\bibitem{selgra}
B. Bakhti, D. Boukari, M. Karbach, P. Maass, and G. M\"uller,
\emph{Density profiles of a self-gravitating lattice gas in one, two, and three
dimensions},
Phys. Rev. E \textbf{97}, 042131 (2018).



\bibitem{spitzer}  {\small L. Spitzer, {\it The dynamics of the interstellar
medium. III. Galactic distribution}, Astrophys. J.  {\bf 95}, 329 (1942)}

\bibitem{camm}  {\small G.L. Camm, {\it Self-gravitating star systems}, Mon.
Not. R. Astron. Soc.   {\bf 110}, 305
(1950)}

\bibitem{rybicki}  {\small G.B. Rybicki, {\it Exact statistical mechanics of a
one-dimensional self-gravitating system}, Astr. Space. Sci. {\bf 14}, 56 (1971)}

\bibitem{kl}  {\small J. Katz, M. Lecar, {\it A note on the stability of
one-dimensional self-gravitating isothermal systems}, Astr. Space. Sci. {\bf
68}, 495 (1980)}

\bibitem{sc}{\small  C. Sire, P.H. Chavanis, {\it Thermodynamics and collapse of
self-gravitating Brownian particles in D dimensions}, Phys. Rev. E {\bf
66}, 046133 (2002)}

\bibitem{cmct}  {\small P.H. Chavanis, 
{\it Critical mass of bacterial populations and critical temperature of
self-gravitating Brownian particles in two dimensions}, Physica A  {\bf 384},
392 (2007)}



\bibitem{cf}  {\small S. Chandrasekhar, E. Fermi, {\it
Problems of gravitational
stability in the presence of a magnetic field}, Astrophys. J.  
{\bf 118}, 116 (1953)}

\bibitem{stodolkiewicz}  {\small J.S. Stodolkiewicz, {\it On the gravitational
instability of some magneto-hydrodynamical systems of astrophysical interest.
Part III}, Acta
Astr. {\bf 13}, 30
(1963)}

\bibitem{ostriker}  {\small J. Ostriker, {\it The equilibrium of polytropic and
isothermal cylinders}, ApJ {\bf 140}, 1056 (1964)}

\bibitem{salzberg}  {\small A.M. Salzberg, {\it Exact statistical thermodynamics
of gravitational interactions in one and two dimensions}, J. Math. Phys.  
{\bf 6}, 158 (1965)}

\bibitem{klb}  {\small J. Katz, D. Lynden-Bell, 
{\it The gravothermal instability in two dimensions}, Mon. Not. R. Astron. Soc.
{\bf
184}, 709 (1978)}

\bibitem{paddy2d}  {\small T. Padmanabhan, {\it Liouville field theory and the
partition function for two-dimensional Newtonian gravity}, Mon. Not. R. Astron.
Soc.   {\bf
253}, 445 (1991)}

\bibitem{aly}  {\small J.J Aly, {\it Thermodynamics of a two-dimensional
self-gravitating system},
Phys. Rev. E {\bf 49}, 3771 (1994)}

\bibitem{ar}  {\small E. Abdalla, M. Reza Rahimi Tabar, {\it Phase transition in
a
self-gravitating planar gas}, Phys. Lett. B {\bf 440}, 339
(1998)}

\bibitem{ap}  {\small J.J Aly, J. Perez, {\it Thermodynamics of a
two-dimensional unbounded self-gravitating system}, Phys. Rev. E {\bf 60}, 5185
(1999)}

\bibitem{virialD}  {\small P.H. Chavanis, {\it Newtonian gravity in $d$
dimensions}, C. R. Physique  {\bf 7}, 331
(2006)}

\bibitem{bppv}  {\small P.H. Chavanis, {\it Virial theorem for rotating
self-gravitating Brownian particles and two-dimensional point vortices}, Int. J.
Mod. Phys. B {\bf 26}, 1241002
(2012)}

\bibitem{aaiso}  {\small P.H. Chavanis, {\it Gravitational instability of finite
isothermal spheres},
Astron. Astrophys. {\bf 381}, 340
(2002)}

\bibitem{poincare}  {\small H. Poincar\'e, {\it Sur l'\'equilibre d'une masse
fluide anim\'ee d'un mouvement de rotation}, Acta Math. {\bf 7}, 259 (1885)}










\end{thebibliography}
\end{document}